\documentclass[usenatbib]{mnras}
\usepackage{graphicx, amssymb, aas_macros, natbib}
\usepackage{booktabs,caption}
\usepackage[flushleft]{threeparttable}
\usepackage{amsmath}
\usepackage{xcolor}
\hypersetup{colorlinks=true, citecolor=blue, urlcolor=blue, linkcolor=blue}

\makeatletter
\newcommand*{\rom}[1]{\expandafter\@slowromancap\romannumeral #1@}
\makeatother

\newcommand{\vmax}{V_{\rm{max}}}

\newcommand{\mvir}{M_{\rm{vir}}}

\newcommand{\tvir}{T_{\rm{vir}}}

\newcommand{\mstar}{M_{\star}}
\newcommand{\msun}{M_{\odot}}

\newcommand{\hmpc}{h^{-1} \, {\rm Mpc}}

\newcommand{\mpc}{{\rm Mpc}}
\newcommand{\kpc}{{\rm kpc}}
\newcommand{\kms}{{\rm km \, s}^{-1}}

\newcommand{\mhalo}{M_{\rm halo}}

\newcommand{\tkBB}[1]{\textcolor{red}{#1}}  

\begin{document}

\title[Sterile Neutrino Dark Matter on FIRE]{Warm FIRE: Simulating Galaxy Formation with Resonant Sterile Neutrino Dark Matter}

\author[B. Bozek et al.] 
{Brandon Bozek$^{1}$\thanks{bozek@astro.as.utexas.edu}, 
Alex Fitts$^1$,
Michael~Boylan-Kolchin$^1$, 
Shea~Garrison-Kimmel$^{2}$, \newauthor 
Kevork~Abazajian$^3$, 
James~S.~Bullock$^3$,
Du\v{s}an~Kere\v{s}$^4$,\newauthor 
Claude-Andr\'{e}~Faucher-Gigu\`{e}re$^5$,
Andrew~Wetzel$^6$,
Robert~Feldmann$^7$,\newauthor 
Philip~F.~Hopkins$^2$\\
$^1$Department of Astronomy, The University of Texas at Austin, 2515 Speedway, Stop C1400, Austin, TX 78712, USA\\
$^2$TAPIR, California Institute of Technology, Pasadena, CA 91125, USA\\
$^3$Department of Physics and Astronomy, University of California,
Irvine, CA 92697, USA\\
$^4$Department of Physics, Center for Astrophysics and Space Sciences, University of California, San Diego, La Jolla, CA, USA\\
$^5$Department of Physics and Astronomy and CIERA, Northwestern University, 2145 Sheridan Road, Evanston, IL 60647, USA\\
$^6$Department of Physics, University of California, Davis, CA 95616, USA\\
$^7$Institute for Computational Science, University of Zurich, Zurich CH-8057, Switzerland}

\pagerange{\pageref{firstpage}--\pageref{lastpage}} 
\pubyear{2015}

  \maketitle
\label{firstpage}

\begin{abstract} 
We study the impact of a warm dark matter (WDM) cosmology on dwarf galaxy formation through a suite of cosmological hydrodynamical zoom-in simulations of $M_{\rm halo} \approx10^{10}\,\msun$ dark matter halos as part of the Feedback in Realistic Environments (FIRE) project. A main focus of this paper is to evaluate the combined effects of dark matter physics and stellar feedback on the well-known small-scale issues found in cold dark matter (CDM) models. We find that the $z=0$ stellar mass of a galaxy is strongly correlated with the central density of its host dark matter halo at the time of formation, $z_{\rm f}$, in both CDM and WDM models. WDM halos follow the same $\mstar(z=0)-\vmax(z_{\rm f})$ relation as in CDM, but they form later, are less centrally dense, and therefore contain galaxies that are less massive than their CDM counterparts. As a result, the impact of baryonic effects on the central gravitational potential is typically diminished relative to CDM. However, the combination of delayed formation in WDM and energy input from stellar feedback results in dark matter profiles with lower overall densities. The WDM galaxies studied here have a wider diversity of star formation histories (SFHs) than the same systems simulated in CDM, and the two lowest $\mstar$ WDM galaxies form all of their stars at late times. The discovery of young ultra-faint dwarf galaxies with no ancient star formation -- which do not exist in our CDM simulations -- would therefore provide evidence in support of WDM. 
\end{abstract}
\tkBB{}

\section{Introduction}
\label{sec:intro} 

The leading class of dark matter particle candidates is phenomenologically ``cold", which is consistent with the weakly interacting massive particle (WIMP) paradigm, axion dark matter, and many other particle physics models. Numerous studies using high-resolution numerical simulations have demonstrated the dark energy + CDM ($\Lambda$CDM) model's ability to reproduce the observed properties of the Universe on scales above $\sim 1~\mpc$ \citep{Frenk2012, Kuhlen2012,Primack2015}. On smaller scales, comparing predictions of the $\Lambda$CDM model from dark-matter-only simulations with observations of low-mass galaxies reveals several issues \citep{Bullock2017}.  Those issues include the over-prediction of low-mass subhalos compared with counts of dwarf galaxies in the Local Group (Missing Satellites Problem -- \citealt{Klypin1999,Moore1999}) and a mismatch of the predicted dark matter content of dark matter halos and the dark matter density inferred from observations of dwarf galaxies expected to reside in those halos (the cusp-core Problem -- \citealt{Flores1994,Moore1994} and the Too Big To Fail (TBTF) Problem -- \citealt{Boylan-Kolchin2011,Boylan-Kolchin2012}). The proposed resolutions of these issues within the context of general relativity appeal to galaxy formation physics altering the predicted dark matter halo properties, new dark matter physics, or some combination of the two effects (for a discussion of these and other issues in the context of modified gravity theories, see \citealt{Famaey2012}). 

Hydrodynamic simulations over a range of scales, from low-mass dwarf galaxies to larger Milky Way-sized galaxies, have demonstrated the importance of baryonic processes in addressing small-scale dark matter issues and building realistic Local Group galaxy populations within the $\Lambda$CDM paradigm \citep{Governato2012,Zolotov2012,Sawala2016,Wetzel2016}. For example, the Missing Satellites problem can be remedied by suppressing star formation in low-mass halos through a combination of stellar feedback and photoionization from a UV background \citep{Bullock2000,Benson2002,Somerville2002,Kravtsov2004}. The discrepancy in counts of luminous satellites and subhalos can be further alleviated by tidal disruption of dark matter subhalos in simulations that include a Galactic disk potential (\citealt{DOnghia2010,Garrison-Kimmel2017,Sawala2017}; though see \citealt{vandenBosch2018}). High-resolution hydrodynamical simulations of the Milky Way-mass halos including these two components have reproduced a satellite stellar mass function that is consistent with observations down to $M_{\star} \gtrsim 10^5 \msun$ \citep{Sawala2016,Wetzel2016}. 

The TBTF and cusp-core problems may be resolved via repeated stellar bursts driving baryonic material from halo centers and triggering gravitational potential fluctuations that reduce central dark matter density \citep{Governato2012,Zolotov2012,Di-Cintio2014,DiCintio2014b,Onorbe2015,Chan2015,Read2016,Brooks2017}. However, the input physics and star formation prescriptions of different simulations have resulted in different predictions for the degree to which dark matter structure is modified by stellar feedback events. For example, while \cite{Sawala2016} and \cite{Wetzel2016} both reproduce many observed properties of Local Group galaxies including resolving the TBTF problem, they disagree on whether the inner density profile of dwarf galaxies feature a core or a cusp. Understanding dwarf galaxy formation and evolution is critical to testing both the nature of dark matter and the coupling between dark matter and galaxy formation physics.

Small-scale issues have motivated consideration of WDM as a compelling alternative to the standard CDM scenario. WDM models feature the same large-scale predictions of the CDM model, but incorporate a non-negligible velocity distribution that erases primordial perturbations with masses below a model-dependent scale \citep{Hogan2000,Sommer-Larsen2001,Bode2001,Barkana2001}. Dissipationless WDM simulations have shown that fewer low-mass halos form as a result thereby addressing the Missing Satellites problem independent of the details of galaxy formation models \citep{Colin2000,Bode2001,Polisensky2011,Lovell2012,Anderhalden2013,Bozek2016,Horiuchi2016}. WDM halos do not suffer from the TBTF problem to the same degree as they form later with a reduced central density than CDM halos of similar masses \citep{Lovell2012,Horiuchi2016,Lovell2017b}. WDM halos can also feature cored dark matter density profiles \citep{Tremaine1979,Dalcanton2001}, but the free-streaming scales of the models we consider here are not expected to resolve the cusp-core problem \citep{Villaescusa-Navarro2011,Maccio2012}. For example, a thermal WDM particle with mass, $m_{\rm thm} = 2\,{\rm keV}$ is predicted to produce a $r \approx 10\,{\rm pc}$ core in a $10^{10}\,\msun$ halo, which falls significantly below the $\sim100-1000\,{\rm pc}$ core size observationally inferred for dwarf galaxies \citep{Gilmore2007,Walker2011,Oh2011}.

In this work, we consider a specific WDM particle model: a resonantly-produced sterile neutrino (RPSN; \citealt{ShiFuller1999}). In addition to resolving small-scale issues, the RPSN model is motivated as a potential source of the significant, but tentative, detection of a $3.55\,\rm{keV}$ line in the X-ray flux observed in the center of the MW, M31, the Perseus cluster, and stacked observations of other clusters \citep{Boyarsky2014, Bulbul2014,Boyarsky2015,Iakubovskyi2015,Abazajian2017}. A RPSN with the proper mixing angles and a mass of $7.1\,\rm{keV}$ can radiatively decay into an active neutrino and a $E=3.55\,\rm{keV}$ photon \citep{ShiFuller1999,Abazajian2014}. While both the existence of the line and its interpretation remain controversial, the possibility that the line represents a detection of dark matter makes it a compelling and intriguing signal. As detailed in Section \ref{sec:Sims}, the parameterization of the RPSN model we adopt is consistent with tentative detections of this line in galaxy and galaxy cluster observations while also providing the largest free-streaming effects allowable by observations of small-scale structure \citep{Bozek2016,Horiuchi2016,Bose2016b}.

A key threshold for galaxy formation within the CDM paradigm is a halo mass of $10^{10}\,\msun$. This scale corresponds to the transition between mass regimes where stellar feedback is effective (at higher masses) to ineffective (at lower masses) in modifying halos' central dark matter distributions  \citep{Governato2012,Di-Cintio2014,Onorbe2015}. Halos at this mass are expected to be susceptible to reionization-induced feedback that reduces their baryonic content and subsequently affects the SFHs of the central galaxies \citep{Efstathiou1992, Hoeft2006, Okamoto2008, Noh2014}. \cite{Fitts2017} studied a suite of halos at this mass scale and demonstrated the combination of stellar feedback and UV-suppression effects provides a diversity of SFHs that range from continuous star formation to early self-quenching to halos that do not form any stars by $z=0$. Consistent with other studies \citep{Di-Cintio2014,Chan2015,Tollet2016}, they also identified a critical threshold in the stellar mass-halo mass ratio of $2\times 10^{-4}$ for $10^{10}\,\msun$ halos: halos above this threshold are able to significantly modify their density profile, while halos below are not. The resulting galaxies are classical dwarf galaxy analogues with clear predictions for the resolution of the TBTF and cusp-core problems within the CDM paradigm.

\begin{table*}
\begin{center}
\caption{Global $z=0$ properties of WDM halos, including comparisons with their CDM counterparts, for the hydrodynamical simulations. {\it Column 1}: Halo name; {\it Column 2}: maximum amplitude of rotation curve; {\it Column 3}: virial mass; {\it Column~4}: stellar mass of central galaxy [defined as $\mstar(<0.1\,R_{\rm vir})$]; {\it Column 5}: formation redshift; {\it Column 6}: WDM-CDM virial mass ratio; {\it Column 7}: WDM-CDM stellar mass ratio; {\it Column 8}: WDM-CDM central density ratio.}  
\label{tab:halolist}
\begin{tabular}{l | c | c | c | c | c | c | c }
\hline
Halo & $\vmax \, [\kms]$ & $M_{\rm vir} \, [10^{10}\msun]$ & $M_{\star} \, [10^{6}\msun]$ & $z_f$ & $M_{\rm vir,WDM}/M_{\rm vir,CDM}$ & $M_{\star,{\rm WDM}}/M_{\star,{\rm CDM}}$ & $\rho_{\rm WDM}/\rho_{\rm CDM}(500\,{\rm pc})$ \\ \hline 
m10b & 27.59 & 0.74 & -- & 1.24 & 0.81 & -- & 0.56 \\
m10c & 26.36 & 0.67 & 0.02 & 2.89 & 0.77 & 0.03 & 0.80 \\
m10d & 28.40 & 0.70 & 0.38 & 5.37 & 0.84 & 0.25 & 0.68 \\
m10e & 28.23 & 0.85 & 0.29 & 2.12 & 0.85 & 0.15 & 0.79 \\
m10f & 31.23 & 0.76 & 2.76 & 5.37 & 0.90 & 0.67 & 0.54 \\
m10h & 35.34 & 1.10 & 2.73 & 4.16 & 0.88 & 0.35 & 0.77 \\
m10k & 36.12 & 1.10 & 4.10 & 3.77 & 0.97 & 0.39 & 0.92 \\
m10m & 35.73 & 1.08 & 2.22 & 5.59 & 0.96 & 0.15 & 0.94 \\
\end{tabular}
\end{center}
\end{table*}

Previous hydrodynamical simulations of dwarf galaxies in the WDM paradigm have focused on halos that are more massive, in a different environment, or have been limited to a single $\mhalo = 10^{10}\msun$ halo and therefore have not explored the range of SFHs that are possible at this mass scale \citep{Governato2015,Colin2015,Gonzalez-Samaniego2016,Lovell2017}. Additionally, previous works have used other galaxy formation prescriptions which can result in different galaxy properties at this mass scale, as discussed above for CDM. A goal of this paper is to explore that galaxy formation threshold in a resonantly produced sterile neutrino WDM model where the halo mass is also near the half-mode mass (see Section \ref{sec:Sims}) where free-streaming effects are significant. We seek to answer several questions: How does the central density profile respond when both stellar feedback and free-streaming effects are prevalent? What happens to the SFHs in galaxies that sit at both dark matter and galaxy formation thresholds? Can galaxy properties in dwarf galaxies that reside in dark matter halos where dark matter effects are so prevalent be used to discriminate between dark matter models? 

To address these questions, we resimulate 8 of the 12 dwarf galaxies from \cite{Fitts2017} in a resonantly-produced sterile neutrino cosmology in order to make a one-to-one comparison between CDM and WDM effects in our simulations. All of our simulations use the GIZMO code\footnote{A public version of GIZMO is available at \url{http://www.tapir.caltech.edu/∼phopkins/Site/GIZMO.html}}\citep{Hopkins2014,Hopkins2017} and the FIRE-2 galaxy formation and feedback model \citep{HopkinsFIRE2017}.\footnote{FIRE project website: \url{http://fire.northwestern.edu}} We provide an overview of our simulations and the sterile neutrino model in Section \ref{sec:Sims}. In Section \ref{sec:Results} we present our results for the WDM halo properties and their central galaxies in a WDM cosmology. We discuss these results in Section \ref{sec:Disc} and conclude in Section \ref{sec:Conc}.

\section{Simulation Details and WDM Structure Formation Theory}
\label{sec:Sims} 

In CDM, structure formation proceeds hierarchically: larger halos are built up through the merging of smaller dark matter halos. As a result of free-streaming effects, WDM structure formation may proceed differently depending on the WDM model parameterization and the halo mass scale under consideration \citep{Bode2001,Barkana2001,Smith2011,Schneider2012}. WDM models define two wavenumbers that are relevant for structure formation, the free-streaming scale ($k_{\rm fs}$) and the half-mode scale ($k_{1/2}$). Below the free-streaming scale, primordial perturbations are erased by the time structure formation begins, wiping out the seeds of low-mass dark matter halos. The half-mode scale is defined implicitly as $T_{\rm rel}(k_{1/2}) = \sqrt{P_{\rm WDM}/P_{\rm CDM}} = 0.5$, where $T_{\rm rel}$ is the `relative' transfer function that describes the  suppression of the WDM model relative to CDM. WDM halos with masses above the half-mode mass, 
\begin{equation}
M_{1/2} = \frac{4 \pi}{3}\,\left(\frac{\pi}{k_{1/2}}\right)^3\,\overline{\rho}\,,
\end{equation}
build up in a similar hierarchical fashion as in CDM \citep{Smith2011,Schneider2012}. Below the half-mode mass but above the free-streaming mass, 
\begin{equation}
M_{\rm fs} = \frac{4 \pi}{3}\,\left(\frac{\pi}{k_{\rm fs}}\right)^3\,\overline{\rho}\,,
\end{equation}
free-streaming effects are significant and the mass assembly history of WDM halos may be far different than their CDM counterparts: hierarchical formation may break down and WDM halos may form through monolithic collapse \citep{Bode2001,Barkana2001,Smith2011,Schneider2013,Angulo2013}.

This work considers sterile neutrinos that are resonantly produced in the presence of a large lepton asymmetry \citep{ShiFuller1999}; the resulting non-thermal momentum distribution depends on a combination of the mixing angle between sterile and active neutrinos, the cosmological lepton number, and the sterile neutrino mass. The model parameter choices set the momentum distribution and defines the shape of the `relative' transfer function. We calculate the exact transfer function using the formulation of \citet{Venumadhav2016} for a $m_{\nu} = 7.1\,{\rm keV}$ sterile neutrino and model parameters that are consistent with the observed $3.55\,{\rm keV}$ X-ray line. Our model choice is then defined by a mixing angle of $\sin^{2}(2\theta) = 2.9 \times 10^{-11}$ (hereafter, S229). As discussed in \cite{Bozek2016}, the thermal root-mean-square velocity of our model is a small fraction of the typical Zel`dovich velocities at the start of the simulation. As the contribution is negligible, we do not include the relic velocity distribution of the WDM in the initial conditions of our simulations.  

The S229 model is selected for free-streaming behavior that addresses both the Missing Satellites and TBTF problems while remaining consistent with the current count of galaxies in the Local Group \citep{Bozek2016,Horiuchi2016}. The TBTF problem is not solved completely by the free-streaming component of the model, which motivates studying this model in full hydrodynamical simulations. The half-mode mass of the S229 model, $M_{1/2} =1.2\times10^9\,\msun$, is comparable with a thermal WDM model with $m_{\rm THM} = 2\,{\rm keV}$, however the S229 model has a smaller free-streaming mass of $M_{\rm fs} \sim 2.5\times10^5\,\msun$ because of its `colder' intrinsic velocity \citep{Venumadhav2016} distribution.\footnote{We will refer to models with smaller (larger) free-streaming lengths relative to S229 as ``colder" (``warmer") models} Given that the final virial mass of the halos we consider here is $M_{\rm vir} = 10^{10}\msun$, the mass assembly history of WDM halos may be far different than their CDM counterparts since they form at a lower mass where these free-streaming effects are relevant. 

Numerical simulations of WDM models suffer from numerical artefacts induced by discreteness noise \citep{Wang2007}; this effect occurs on scales with essentially no power in the initial conditions and is therefore limited to mass scales below $M_{\rm lim} = 10.1\, \bar{\rho} \, d\,k_{\rm peak}^{-2}$ \citep{Wang2007}. Our $z=0$ halos should be insensitive to these numerical effects, as they have masses significantly in excess of $M_{\rm lim} = 2.4\times10^{7} \,\msun$ in our simulations (where we have followed \cite{Lovell2014} and multiplied the \cite{Wang2007} definition by $\kappa = 0.5$). We note that while we do not remove artificial halos from our analysis, contamination from spurious halos does not appear to impact the assembly histories of our dark matter halos or their central galaxies: the main features of our WDM simulations track CDM histories, as detailed below.

We selected eight halos from the CDM simulation suite of \citet{Fitts2017} to be resimulated in WDM; we simulated dark-matter-only (DMO) and hydrodynamical versions in each case. The halos were selected for zoom-in simulations from a parent box of $L=25\, \hmpc$; the targeted halo was required to be isolated from any more massive halo by at least 3 virial radii (of the more massive halo), allowing us to study the formation of the halos and the galaxies independent of environmental effects. Additional details on halo selection can be found in \citet{Fitts2017}. 

Our simulations use the FIRE-2 galaxy formation and feedback model with identical physics and model parameters as those in \cite{HopkinsFIRE2017}. We briefly summarize the FIRE-2 model here; a more detailed description can be found in \cite{HopkinsFIRE2017}. Gas cooling is computed for ionized, atomic, and molecular gas from $T\sim10-10^{10}{\rm K}$ and includes metal-line cooling for 11 elements. Heating and ionization is incorporated from (1) local stellar sources and (2) a redshift-dependent, spatially uniform meta-galactic UV background (updated from \citealt{Faucher-Giguere2009})\footnote{This UV background model is available at \url{http://galaxies.northwestern.edu/uvb}}. Star formation proceeds in locally self-gravitating, dense, self-shielding molecular, Jeans-unstable gas. Feedback sources have inputs taken from stellar evolution models and include: SNe I and II, stellar mass loss from O/B-star and AGB-winds, and multi-wavelength photo-heating and radiation pressure. The gas particle mass is $500\,\msun$ and the dark matter particle mass is $2500\,\msun$, with physical force resolution of $h_{\rm b} = 2\,\rm{pc}$ and $\epsilon_{\rm DM} = 35\,\rm{pc}$. Simulations are run in a WMAP7 cosmology: $\Omega_{\rm m} = 0.266, \,\sigma_{8} = 0.801, \,\Omega_{\Lambda} = 0.734, \,n_{\rm s} = 0.963, \text{ and } \rm{h} = 0.71$ \citep{Larson2011}. Initial conditions for the simulations are created with the MUSIC code \citep{Hahn2011}. 

We use the Amiga Halo Finder (AHF; \citealt{Knollmann2009}) to identify self-bound dark matter halos. Following the analysis outlined in \citet{Fitts2017}, we use the iterative ``shrinking spheres" centering routine \citep{Power2003} to identify halo centers. For DMO simulations, we correct particle masses using $m_{\rm p} \rightarrow (1-f_{\rm b})\,m_{\rm p}$, where $f_{\rm b} = \Omega_{\rm b}/\Omega_{\rm m}$ is the cosmic baryon fraction, which effectively mimics maximal baryonic mass loss for DMO runs. All virial quantities are defined according to \citet{Bryan1998}. The virial overdensity relative to $\rho_{\rm c}$ is $\Delta_{\rm c} = 96.5$ for our chosen cosmology.

\section{Results}
\label{sec:Results}
\subsection{Warm Dark Matter Halo Properties and Assembly in DMO Simulations}
\label{sec:DMO}
\begin{figure}
\begin{center}
\includegraphics[width=0.49\textwidth]{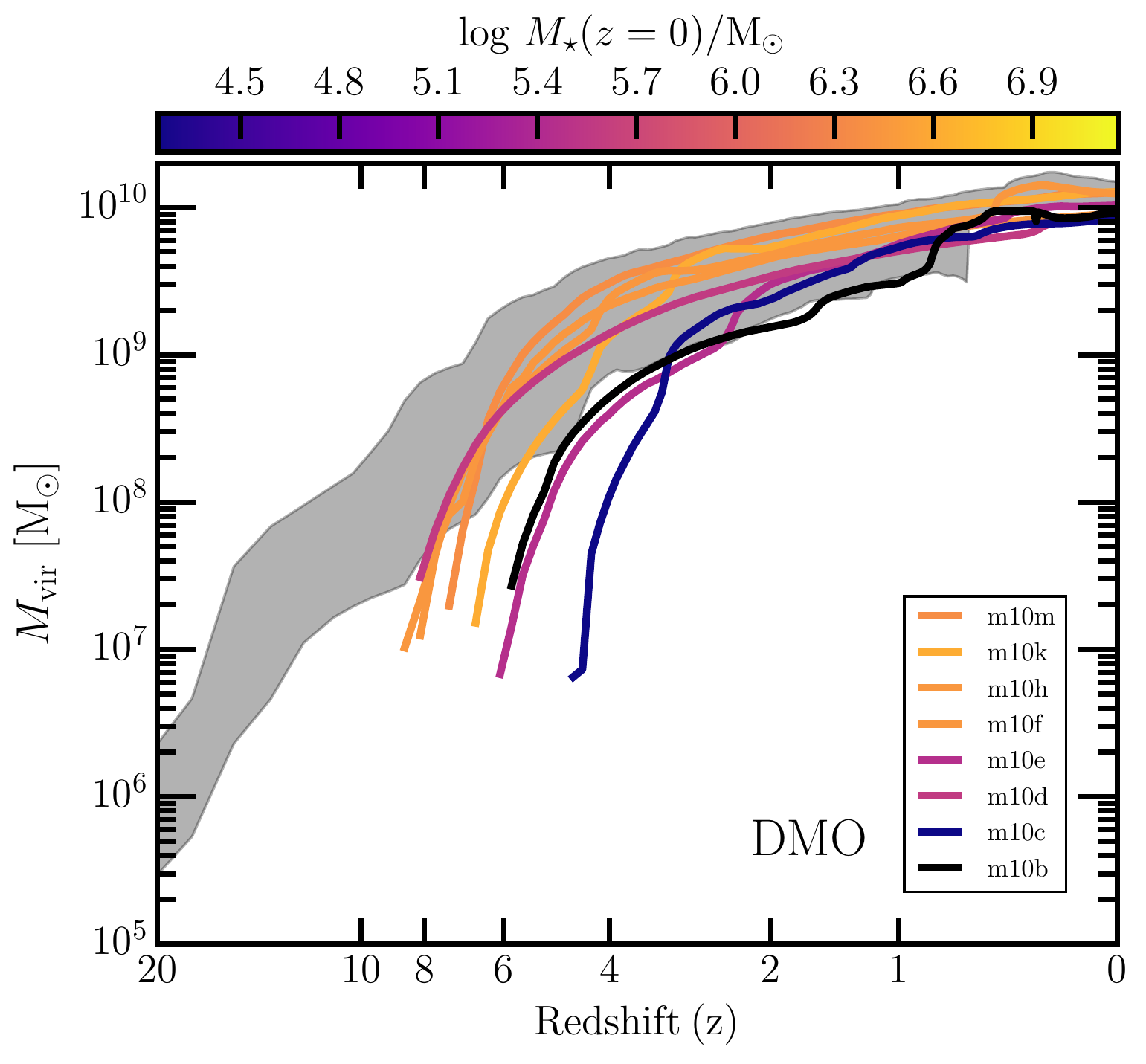}
\caption{The assembly history of the WDM halos (curves) in the DMO simulations. The range of assembly histories for the CDM counterparts of the WDM halos are shown by the gray band. WDM halos collapse later and collapse with a larger mass than their CDM counterparts. The color of the curves corresponds to the $z=0$ stellar mass of the galaxy that resides in each halo in the corresponding hydrodynamical simulations (described further in Sec~\ref{sec:Hydro}). $\mstar(z=0)$ is correlated with the halo mass at the end of the rapid collapse phase.}
\label{fig:mvir_assembly_dmo}
\end{center}
\end{figure}

\begin{figure}
\begin{center}
\includegraphics[width=0.49\textwidth]{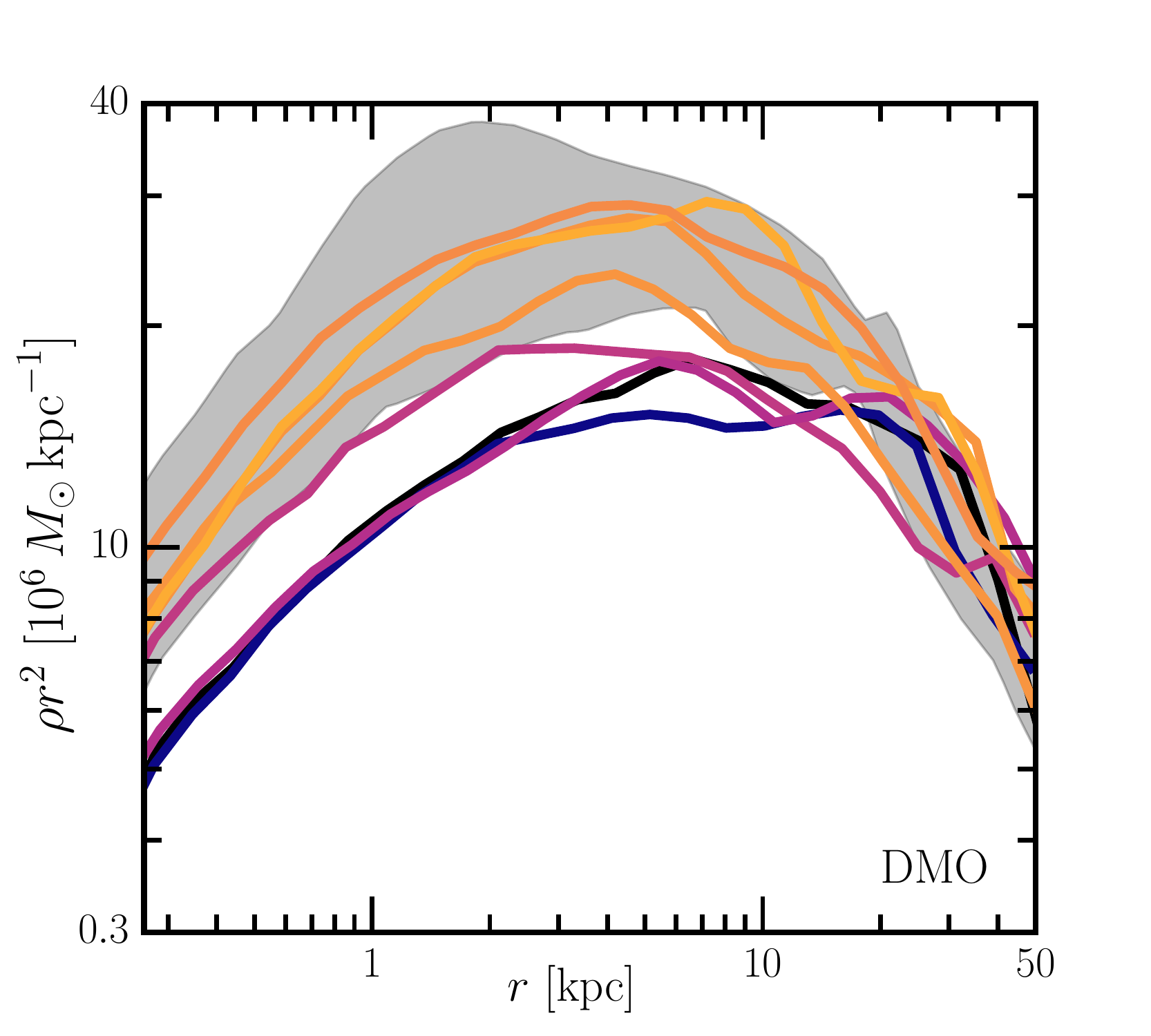} 
\includegraphics[width=0.49\textwidth]{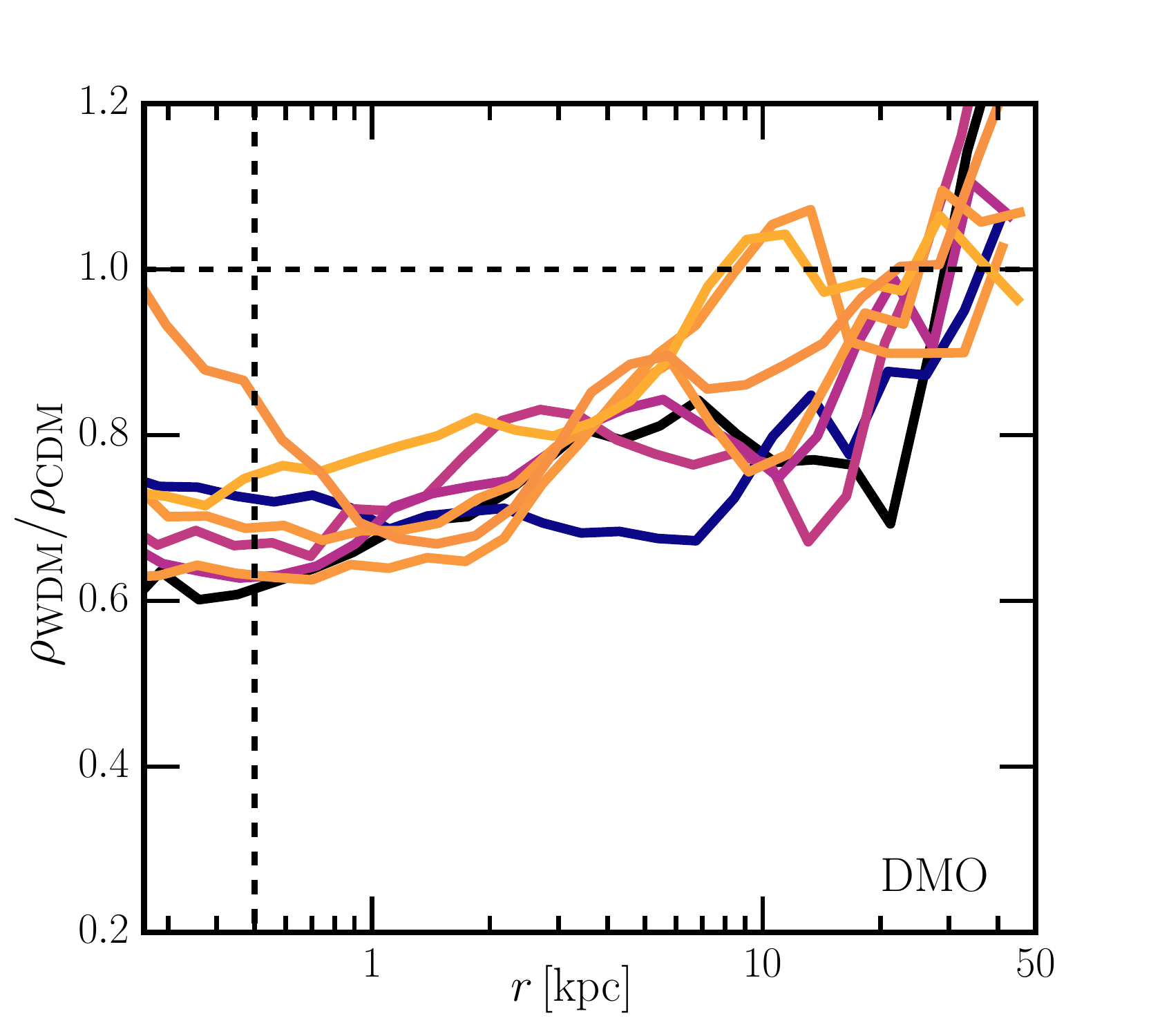}
\caption{\textit{Top Panel:} The density profiles of the WDM halos (colored curves) from the DMO simulations and their CDM (DMO) counterparts (grey band). The line colors are the same as in Figure~\ref{fig:mvir_assembly_dmo}. The downward shift of the curves relative to the grey band indicates the WDM halos are less dense than their CDM counterparts out to $10\,\kpc$, where the curves and grey band overlap. To first order, the stellar mass of the galaxy at $z = 0$ correlates with the central density of the halo, i.e., denser DMO halos have brighter galaxies in their hydrodynamical counterparts. \textit{Bottom Panel:} The ratio of WDM to CDM density profiles from the top panel. There is a $15-40\%$ reduction in the central density of WDM halos relative to CDM at $500\,\rm{pc}$ (dashed vertical line) and the ratios remains below $0.8$ out to a few kpc for all WDM halos; since the figure compares DMO profiles, this reduction is entirely attributable to the effects of free-streaming in WDM. The up-sloping density profile ratio for $r<1\,\rm{kpc}$ for Halo m10m is likely due to a late merger. The degree of the WDM halo density reduction in DMO simulations is independent of the halo's central density.}
\label{fig:rhorsq}
\end{center}
\end{figure}

\begin{figure}
\begin{center}
\includegraphics[width=0.49\textwidth]{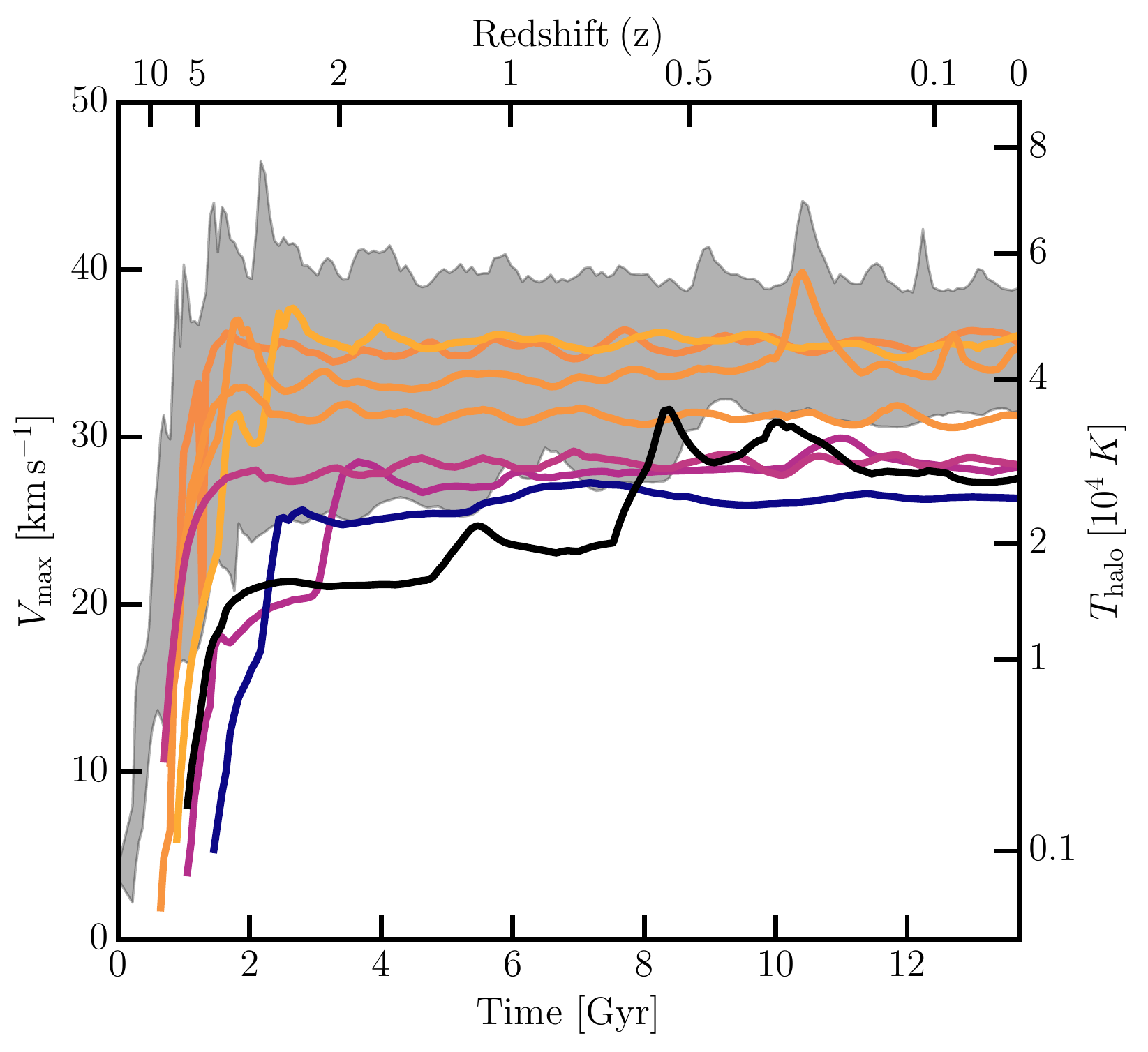}
\caption{$\vmax$ as a function of time for the WDM halos (curves) and their CDM counterparts (grey band) for the hydrodynamical simulations. The equivalent virial temperature, $T_{\mathrm{vir}}$, is given on the right axis. Following the period of rapid collapse, the peak $\vmax$ sets in at $t_{\rm f}\sim1-4\,{\rm Gyr}$ ($z_{\rm f}=5.8-1.7$) and remains mostly constant to $z=0$, with the exception of a late-time major merger for Halo m10b (black). The line color is the same as in previous figures and indicates that $\mstar(z=0)$ is closely connected with the dark matter halo central potential as measured by $\vmax(z_{\rm f})$. Halo m10b forms no stars, as it has a $T_{\mathrm{vir}} < 2\times10^4\,{\rm K}$ for most of its history. The central potential in WDM is reduced relative to CDM in each case.}
\label{fig:VmaxFunc}
\end{center}
\end{figure}

\begin{figure*}
\begin{center}
\includegraphics[width=0.49\textwidth]{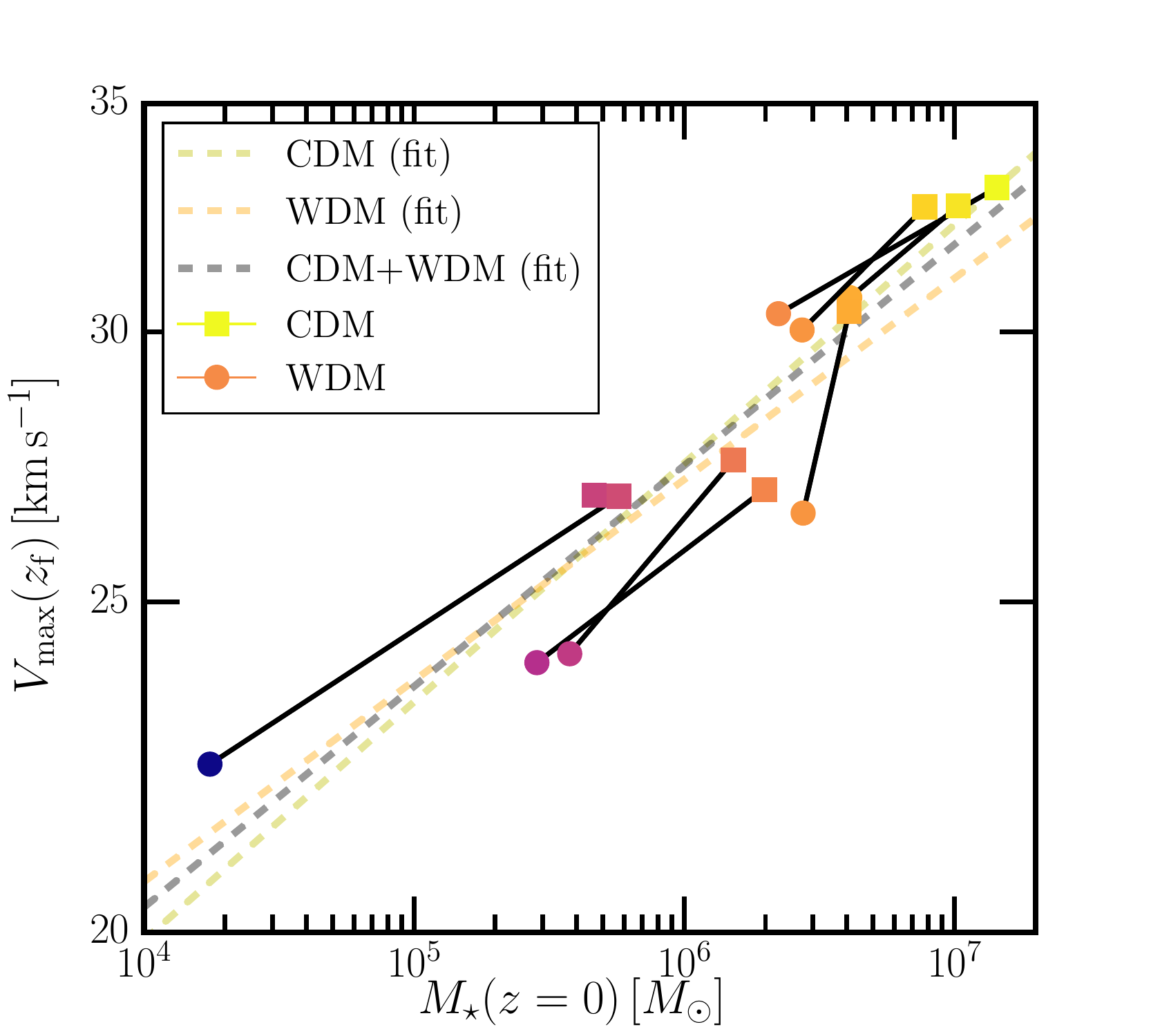}
\includegraphics[width=0.49\textwidth]{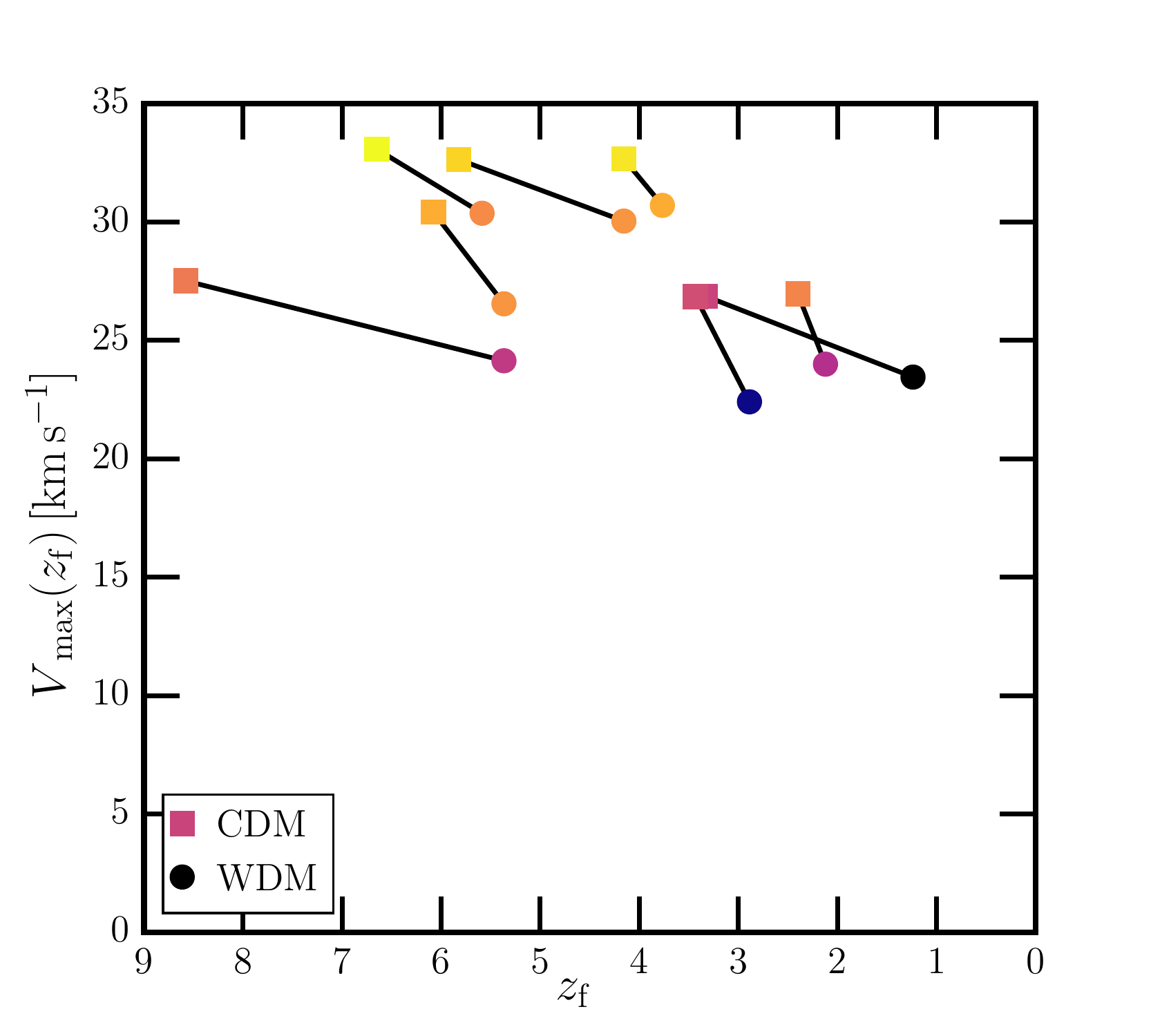}
\caption{\textit{Left Panel:} $\vmax (z_{\rm f})$ at the time of halo formation correlates with the stellar mass of the galaxy at $z = 0$. WDM and CDM halo counterparts are connected by black segments. WDM halos (circles) have a smaller $\vmax (z_{\rm f})$ than their CDM counterparts (squares) and, as a result, a lower stellar mass that moves the galaxy along the $\vmax (z_{\rm f})$-$\mstar (z=0)$ relation. The three $\vmax (z_{\rm f})$-$\mstar (z=0)$ fits shown for WDM halos only (dashed-orange), CDM halos only (dashed-yellow), and WDM and CDM halos together (dashed-grey) are similar. \textit{Right Panel:} The $\vmax (z_{\rm f})$ as a function of formation time. Halos form later and with a $\vmax$ that is reduced by $\sim 10\%$ in WDM relative to CDM. $\vmax (z_{\rm f})$ and $z_{\rm f}$ are not strongly correlated. The color scheme is the same as is used in previous figures.}
\label{fig:VmaxMstar}
\end{center}
\end{figure*}

This section focuses on the assembly histories and halo properties of WDM halos in the DMO simulations. We will highlight how WDM structure formation proceeds differently than in CDM and identify dark matter halo features that will impact galaxy formation when baryons are added later in the hydrodynamical runs. 

Figure \ref{fig:mvir_assembly_dmo} shows the assembly histories of the WDM halos (and their CDM counterparts) in the DMO simulations. The WDM and CDM halos have similar final masses of $M(z = 0)\approx10^{10} \,\msun$, but WDM halos collapse later and the first halos to form are much more massive in WDM than in CDM. We define the collapse mass as the halo mass at the snapshot time when the halo is first identified by AHF as a bound object with a minimum of 200 dark matter particles, corresponding to a mass above $\mvir = 5\times10^5\,\msun$. The collapse masses of the WDM halos all lie between the half-mode mass and free-streaming mass of $M_{\rm fs} \sim 2.5\times10^5\,\msun$ and $M_{1/2} = 1.2\times10^9\,\msun$. The assembly histories of the WDM halos quickly catch up to their CDM counterparts by $z\sim3-4$ and subsequently track the CDM halo growth. The inflection point in the assembly history marks the end of the rapid collapse phase \citep{Wechsler2002,Zhao2003}, which both WDM and CDM halos exit at similar times despite very different initial halo collapse times. The larger collapse mass of the WDM halo allows for a rapid collapse period that is shorter than in CDM. The color of the curves in Figure \ref{fig:mvir_assembly_dmo} correspond to the stellar mass of the galaxy from the hydrodynamical simulation of that halo. The DMO simulation halo masses at the end of the rapid collapse phase ($z \sim 3-4$) in both WDM and CDM models are correlated with the stellar mass of the galaxies that will form in those halos in the hydrodynamical simulations.

Free-streaming smooths density fluctuations on small scales in WDM, resulting in later collapse of WDM halos when the background density of the Universe is lower; this results in a reduced central density relative to CDM. The top panel of Figure \ref{fig:rhorsq} shows the density profiles of the WDM halos at $z=0$ for the DMO simulations; the range of their CDM counterparts is shown in the grey shaded region. The overall reduction in the central density can be seen in the shift of the WDM curves downward from the grey band that extends out to $r\sim 20\,\kpc$; the concentrations of the WDM halos are also reduced, as the peak of $\rho\,r^2$ shifts to larger values relative to the CDM case. The color of the curves in Figure \ref{fig:rhorsq} are the same as in Figure \ref{fig:mvir_assembly_dmo}. The central density of the WDM halos in the DMO simulations are correlated with stellar mass of the galaxies that will form in the corresponding hydrodynamical simulation halos, as was shown in \citet{Fitts2017} for CDM. The denser the WDM halo in the DMO simulation the larger the galaxy stellar mass will be in a hydrodynamical simulation.

The ratio of the density profiles (WDM to CDM) is shown in the bottom panel of Figure \ref{fig:rhorsq}. At $r= 500\,{\rm pc}$\footnote{We make several comparisons of between WDM and CDM at $r=500\,{\rm pc}$ throughout the paper. This value was chosen to represent a fixed point near the center of the halo that is close to the average half-light radius of galaxies in WDM halos and is roughly $2\,r_{\rm conv}$. The convergence radius for each halo is determined by the \cite{Power2003} criterion. We select the largest value of $r_{\rm conv,max}=265\,{\rm pc}$ as the minimum radius for all halos.}, the density ratio of WDM to CDM halos is $0.6-0.85$, which is very similar to the value found by \citet{Bozek2016} for field halos at this mass scale in the vicinity of a simulated Local Group analog. 

Moving out from the center of the halos, the density ratio remains below $0.8$ for the central few kiloparsecs and below unity out to $r\sim10-30\,\kpc$ from the halo centers. Most halos have a relatively constant WDM-to-CDM density ratio out to $r\sim10\,\kpc$. Halo m10m is the one exception: it has a ratio that rises towards the center at $r < 1\,\kpc$, which is the result of a late-time minor merger in the WDM run. 

\subsection{Dark Matter Halo - Stellar Mass Connection}
\label{sec:Hydro}

The previous section focused on the WDM halo properties from DMO simulations. In this section, we consider the WDM halos from the hydrodynamical simulations to explore the connection between the dark matter halo properties and the central galaxy stellar mass. The global properties of our simulated galaxies and their host dark matter halos are detailed in Table \ref{tab:halolist}. 

Figure \ref{fig:VmaxFunc} shows the WDM and CDM halos' $\vmax$ (as colored lines and a grey shaded region, respectively) as a function of time in the hydrodynamical simulations. The peak $\vmax$ value for each WDM halo is lower than its CDM counterpart, which corresponds to the reduction of the central density of the WDM halos shown in Figure \ref{fig:rhorsq} for the DMO simulations. The inflection point in the evolution of $\vmax$ with time between $t = 1-4\,{\rm Gyr}$ ($z_{\rm f}=5.8-1.7$) marks the point where the central halo potential, as measured by $\vmax$, is set. Beyond this point, $\vmax$ is relatively constant for each halo, indicating that subsequent mass accretion does not add to the central gravitational potential. The inflection point in $\vmax$ also coincides with the end of the rapid collapse phase of the halo's mass assembly history as shown in Figure \ref{fig:mvir_assembly_dmo} for the DMO simulations. The WDM halos are therefore being built up inside-out, similar to CDM halos \citep{Diemand2007}, where the inner halo forms first during the rapid collapse phase and the outer layers are added as the halo accretes additional mass. This indicates that WDM halos are growing hierarchically at this mass scale, albeit with a near monolithic inner halo assembly history that differs from CDM.

The color of the curves in Figure \ref{fig:VmaxFunc} are the same as in the previous two figures. As was shown by \citet{Fitts2017} for CDM, the peak $\vmax$ value for each WDM halo is correlated with $\mstar(z=0)$ of each galaxy. One halo, m10b (black curves in Figures \ref{fig:mvir_assembly_dmo}-\ref{fig:VmaxFunc}), does not form any stars. The right y-axis of Figure \ref{fig:VmaxFunc} shows the equivalent virial temperature, defined as ${\rm k}T_{\mathrm{vir}} = 0.5\,\mu\,m_{\rm p}\vmax^2$ (where $\mu=0.59$ for primordial ionized gas and $m_{\rm p}$ is the proton mass), of Halo m10b is below the temperature of reionization heated intergalactic gas ($T\approx2\times10^4\,{\rm K}$) for most of its assembly history preventing the onset of star formation. The star formation histories of the WDM galaxies are discussed in detail in Section \ref{sec:Gals}.

Figure \ref{fig:VmaxMstar} shows the relationship between the formation time of the WDM and CDM halos and the maximum circular velocity at that time. We define the formation time of the halo as the cosmic time where the $\vmax$ function initially reaches a value of $0.85\, \vmax(z=0)$. This choice most accurately selects the critical point in the halo's assembly history where the central halo potential is set. This definition of formation time is similar to that in \citet{Diemand2007}; however, our definition is based on the value of $\vmax$ today instead of the peak value over all time. There are multiple reasons for making this choice. Several halos in our simulations feature recent mergers that result in temporary spikes in the $\vmax$ function, as can be seen in Figure \ref{fig:VmaxFunc}, which would artificially shift formation times to later times if the peak $\vmax$ value was used. Additionally, the halos in our simulations are isolated and not subject to the same late-time environmental effects of \citet{Diemand2007}. Other choices for defining the formation time in the literature that are based on a fraction of the halo's final $M_{\rm vir}$ \citep{Wechsler2002,Gao2005} are less appropriate here, as they occur after the physical mass assembly of the halo \citep{Diemand2007}. 

\begin{figure*}
\begin{center}
\includegraphics[width=0.49\textwidth]{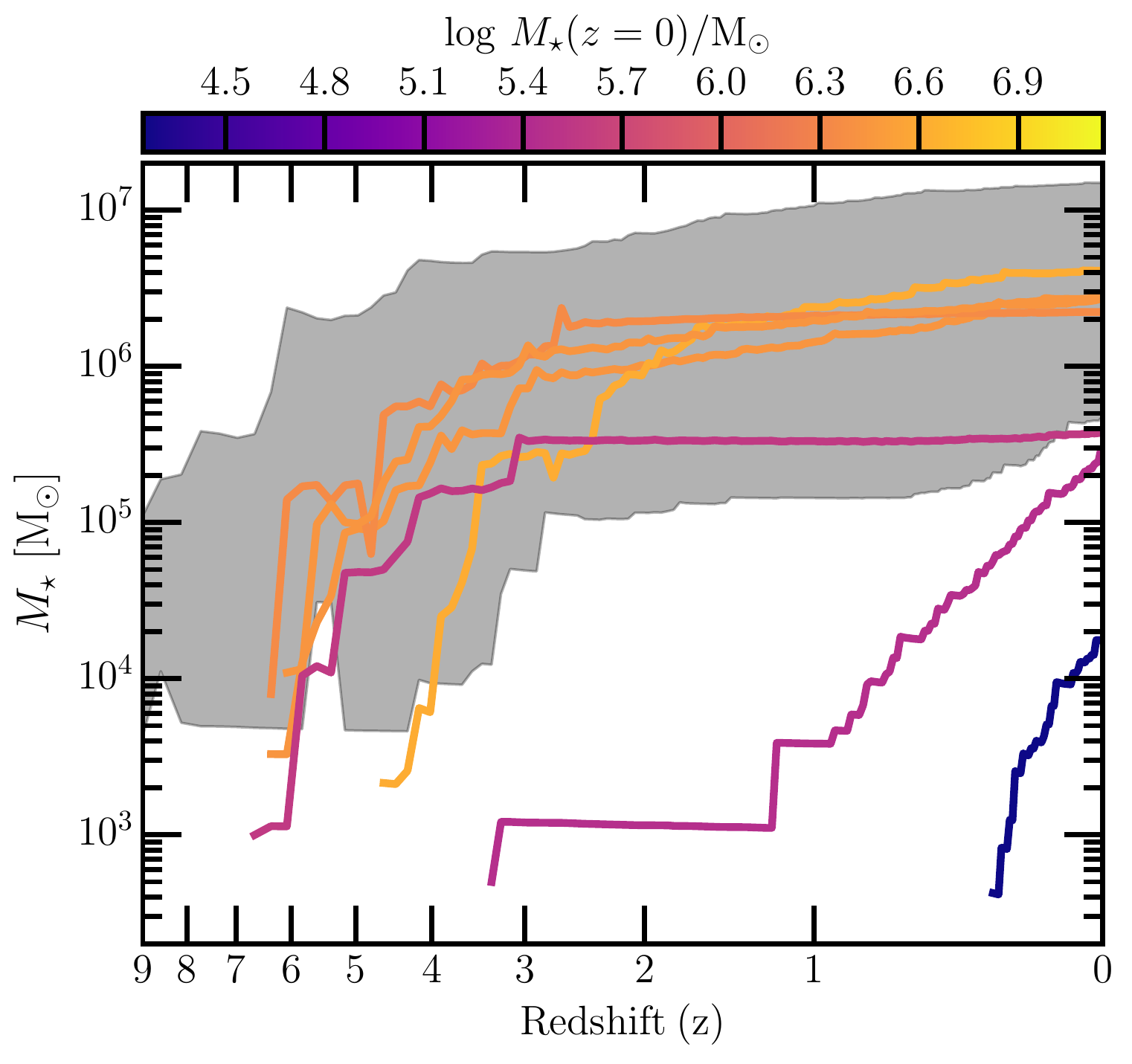}
\includegraphics[width=0.49\textwidth]{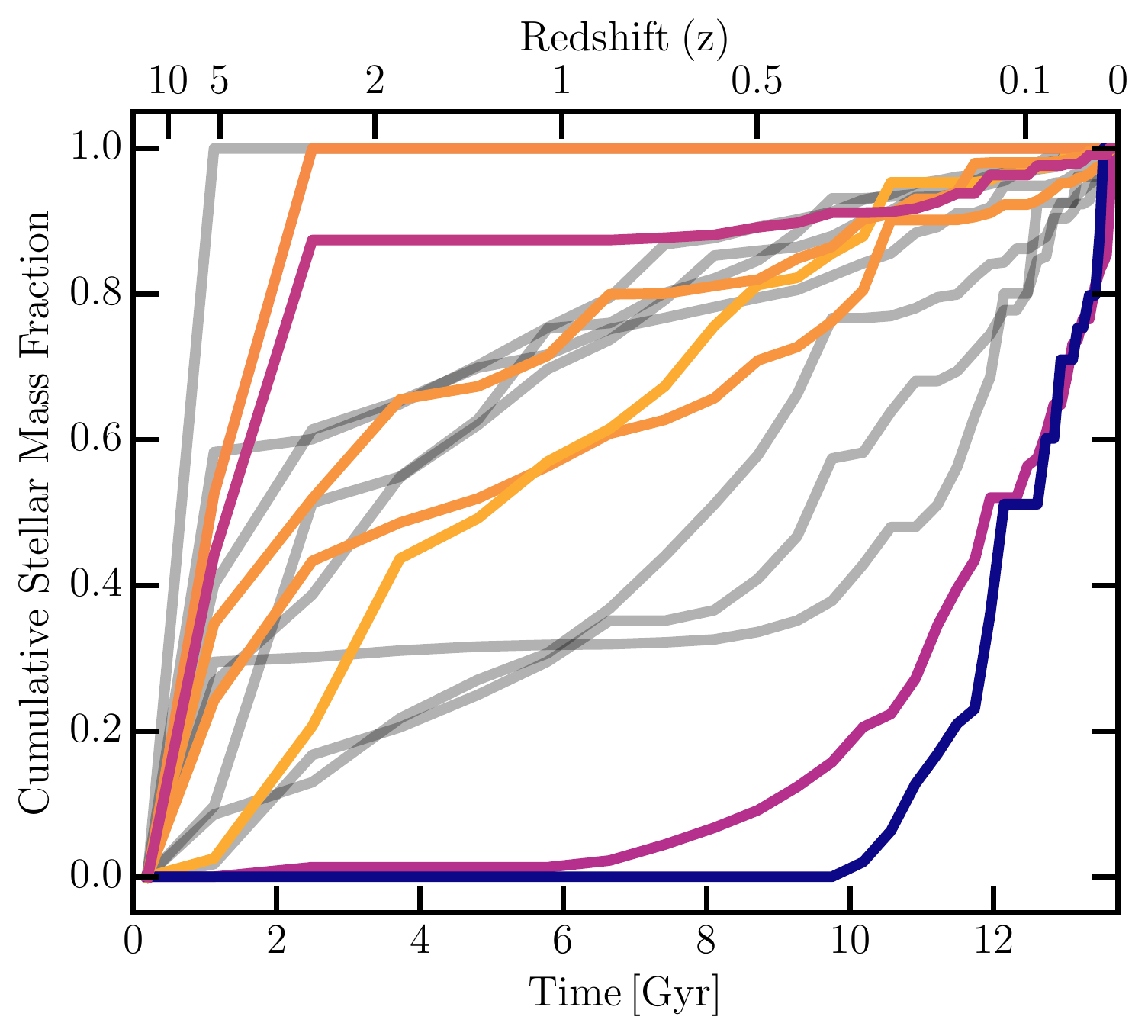}
\caption{\textit{Left Panel:} The SFHs of the galaxies in WDM halos (curves) and the CDM counterparts (grey band). The onset of star formation in WDM galaxies is delayed relative to CDM and the $z=0$ stellar masses are reduced. 
\textit{Right panel:} The cumulative SFHs of the WDM galaxies (colored curves) measured in an ``archaeological" manner (by measuring the birth times of stars in the galaxy at $z=0$). Many of the galaxies have SFHs that are very similar to their CDM counterparts (grey curves), which themselves agree well with observations \citep{Fitts2017}. Two WDM galaxies experience prolonged quenching events. The two least massive galaxies form over $90\%$ of stars after $z=1$; none of the CDM galaxies exhibit this behavior.}
\label{fig:Mvir_Mstar}
\end{center}
\end{figure*}

The $\vmax$ value of a WDM halo at the time of formation correlates strongly with the stellar mass of the galaxy at redshift zero, $\mstar(z=0)$, as shown in the left panel of Figure \ref{fig:VmaxMstar}. The WDM halos follow a similar $\vmax(z_{\rm f})-\mstar(z=0)$ relation as the CDM halos; the decrease in the stellar mass of the galaxies in WDM halos is in proportion to the reduction in the $\vmax(z_{\rm f})$. The dashed grey curve in Figure \ref{fig:VmaxMstar} shows the fit to the $\vmax(z_{\rm f})-\mstar(z=0)$ relation for both WDM and CDM halos:
\begin{equation}
\vmax(z_{\rm f}) = 27.5\,\kms \, \left(\frac{\mstar(z=0)}{10^6\msun}\right)^{0.065}\,.
\end{equation}
The depth of the central WDM halo potential that is set at the end of the rapid collapse phase determines the final stellar mass of the galaxy, as was found to be the case in CDM by \cite{Fitts2017}. 

The left panel of Figure \ref{fig:VmaxMstar} shows that the relation between $\vmax(z_{\rm f})$ and $\mstar(z=0)$ holds across WDM and CDM halos, which is a result of the similar degree of central density suppression in each WDM halo, as shown in the bottom panel of Figure \ref{fig:rhorsq}. Each WDM halo features a $\sim 30\%$ density reduction in the inner $1\,{\rm kpc}$ relative to CDM independent of the overall density of the CDM halo. Combined with a slight decrease in the halo concentration, this effect results in a $\sim10\%$ reduction in the WDM halo's $\vmax$ relative to CDM and fits the predicted $V_{\rm max,WDM}-V_{\rm max,CDM}$ scaling from \citet{Bozek2016}. The $\vmax(z_{\rm f})-\mstar(z=0)$ and $V_{\rm max,WDM}-V_{\rm max,CDM}$ relations could, in principle, be used to predict the WDM halo stellar mass from corresponding CDM simulations. We note that the $\vmax(z_{\rm f})-\mstar(z=0)$ relation has been determined for halos at $10^{10}\,\msun$ and may not hold at higher masses. Another caveat to this prescription are cases such as Halo m10b that does not form stars in WDM. We explore WDM galaxy formation histories in the following sections.

\subsection{Galaxy Properties in WDM}
\label{sec:Gals}

The left-hand panel in Figure \ref{fig:Mvir_Mstar} shows the SFHs of the galaxies residing in the WDM halos. The stellar mass growth histories are varied in detail and span a wider range of assembly tracks than their CDM counterparts, as can be seen by comparing the colored lines (WDM) to the grey band. In the previous section, we established the halo's $\vmax$ at the time of formation as the main determinant of the galaxy's stellar mass. Dividing the WDM galaxies into two groups, high stellar mass ($M_{\star} \geq 10^6\,\msun$) and low stellar mass ($M_{\star} < 10^6\,\msun$), we see that the high stellar mass galaxies all have a $\vmax(z_{\rm f}) > 25\,\kms$ while lower $\vmax(z_{\rm f})$ halos form significantly fewer stars or do not form stars at all. 

Generally, the high $\mstar$ galaxies of our suite reside in halos that collapse earlier ($z_{\rm c} > 6$), have earlier formation times ($z_{\rm f} > 3.5$), and begin to form stars early on in their assembly history ($z_{\star,{\rm i}} > 4$). Earlier formation allows this subset of WDM halos to build a reservoir of gas that can drive sustained star formation. However, these parameters are not exactly correlated with the final stellar mass of the galaxy. For example, the most massive WDM galaxy (Halo m10k) collapses last, has the latest formation time, and does not begin to form stars until $z\sim4.5$ (last amongst the larger stellar mass galaxy group). The onset of star formation in Halo m10k begins even later than in Halo m10d ($z\sim7$), which forms one-tenth as many stars. 

The right panel of Figure \ref{fig:Mvir_Mstar} shows the SFHs as they would be constructed observationally via the method of ``stellar archaeology". Qualitatively, the SFHs of the massive WDM galaxies (as well as Halo m10d from the low-mass group) are comparable with those of CDM galaxies in Fitts et al.~(\citeyear{Fitts2017}, shown in grey in the right panel of Figure \ref{fig:Mvir_Mstar}) and observed dwarf galaxies in the field \citep{Cole2014,Skillman2014}, demonstrating either sustained star formation over their lifetimes or quenching following an early period of star formation. 

However, the SFHs of individual halos in CDM and WDM can be very different. For example, in WDM, Halo m10m has a strong feedback event at $z\sim3$ that drives out all cold interstellar medium (ISM) gas, heats the circumgalactic medium\footnote{We define the ISM to be the gas with $r \leq 0.1R_{\rm vir}$ and the circumgalactic medium to be gas with $r_{\rm ISM} < r \leq R_{\rm vir}$.} to temperatures above the halo virial temperature, and prevents gas cooling onto the galaxy until $z\sim0$. In CDM, Halo m10m does not quench, but instead has a steady growth over its lifetime and is the most massive galaxy in our CDM galaxy sample. Stemming from the delay in WDM structure formation, the start of star formation for all our WDM galaxies is delayed relative to CDM, as was noted by \citet{Governato2015} for a halo at this mass scale simulated with a different set of galaxy formation models. We find that the delay in the onset of star formation is often as short as $\sim0.5\,{\rm Gyr}$ in our simulations; \citet{Governato2015} found a 1-2 Gyr delay in their WDM simulation, consistent with the range of delays found in our suite.

\begin{figure}
\begin{center}
\includegraphics[width=0.49\textwidth]{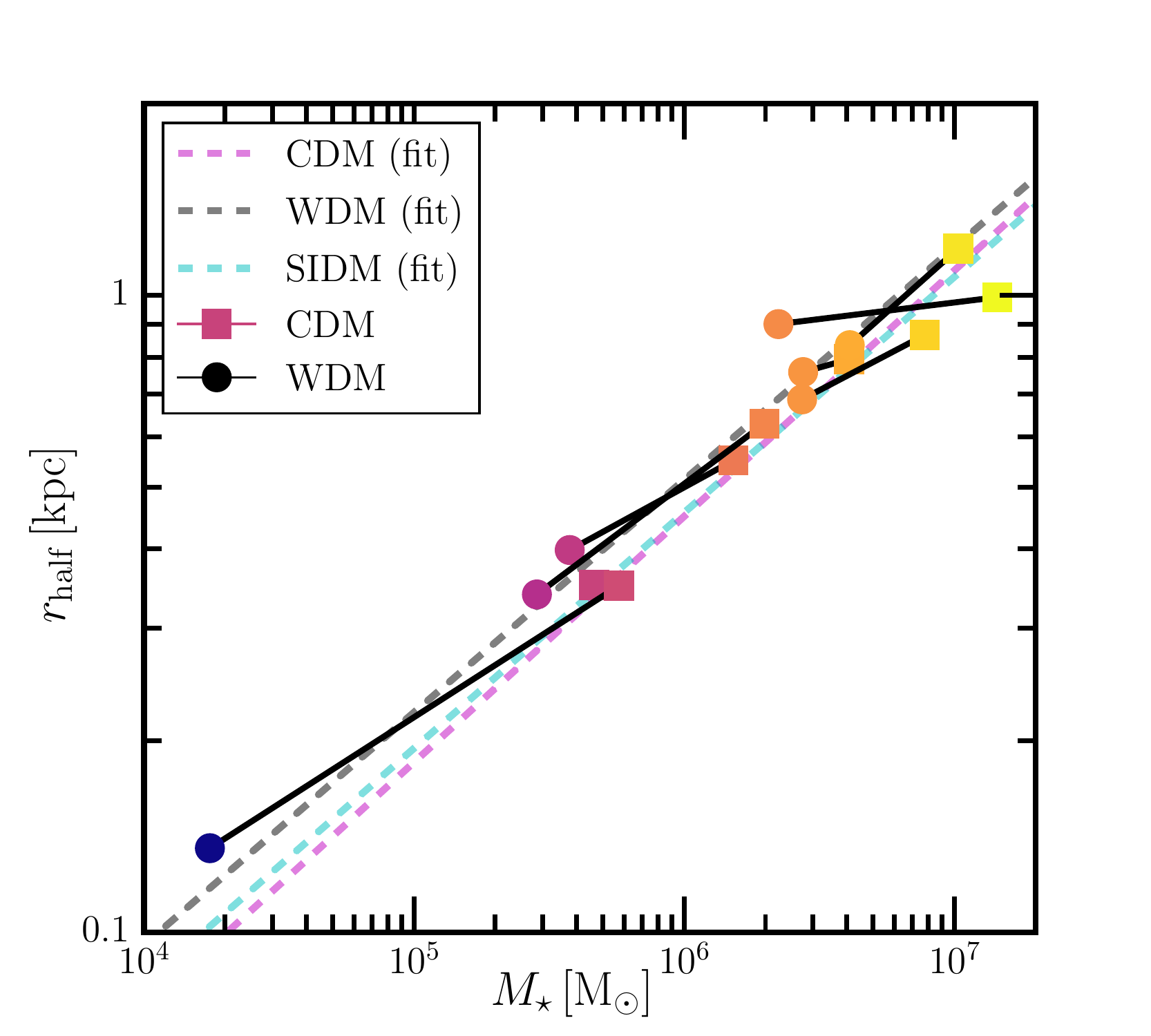}
\caption{The relationship between the stellar half-mass radius and the stellar mass for the WDM and CDM galaxies at $z=0$. While the WDM halos (circles) form fewer stars than their CDM counterparts (squares), the resulting galaxies lie on the same $r_{1/2}-\mstar$ relationship. This relationship is also obeyed by self-interacting dark matter simulations of the same halos (as described in \citealt{Robles2017} and shown in cyan).} 
\label{fig:size_relations}
\end{center}
\end{figure}

The galaxies in the low $\mstar$ group are strongly affected by reionization and stellar feedback and  form the majority of their stars very early or very late (or not at all). The collapse time, formation time, and subsequently the onset of star formation, strongly affect the diversity of SFHs in low $\mstar$ galaxies shown in Figure \ref{fig:Mvir_Mstar}. For example, Halo m10d is one of the earliest collapsing halos in WDM that has an early onset of star formation before quenching; its SFH in WDM qualitatively resembles the CDM version, where it forms all of its stars prior to $z=5$ and remains quenched until $z = 0$. In WDM, however, the quenching is transient: it forms over $80\%$ of its stars before quenching at $z\sim3$ and subsequently restarting star formation at $z\sim 0.5$. The quenching event of WDM Halo m10d is less extreme than in CDM (or compared with WDM Halo m10m) owing to a smaller initial star formation rate that produces less energetic feedback. Unlike Halo m10m, Halo m10d is immediately able to re-accrete warm gas (defined here as $10^4 \,{\rm K} < T < 10^5 \,{\rm K}$), allowing sufficient time to restart star formation by $z=0.5$.

\begin{figure*}
\begin{center}
\includegraphics[width=0.49\textwidth]{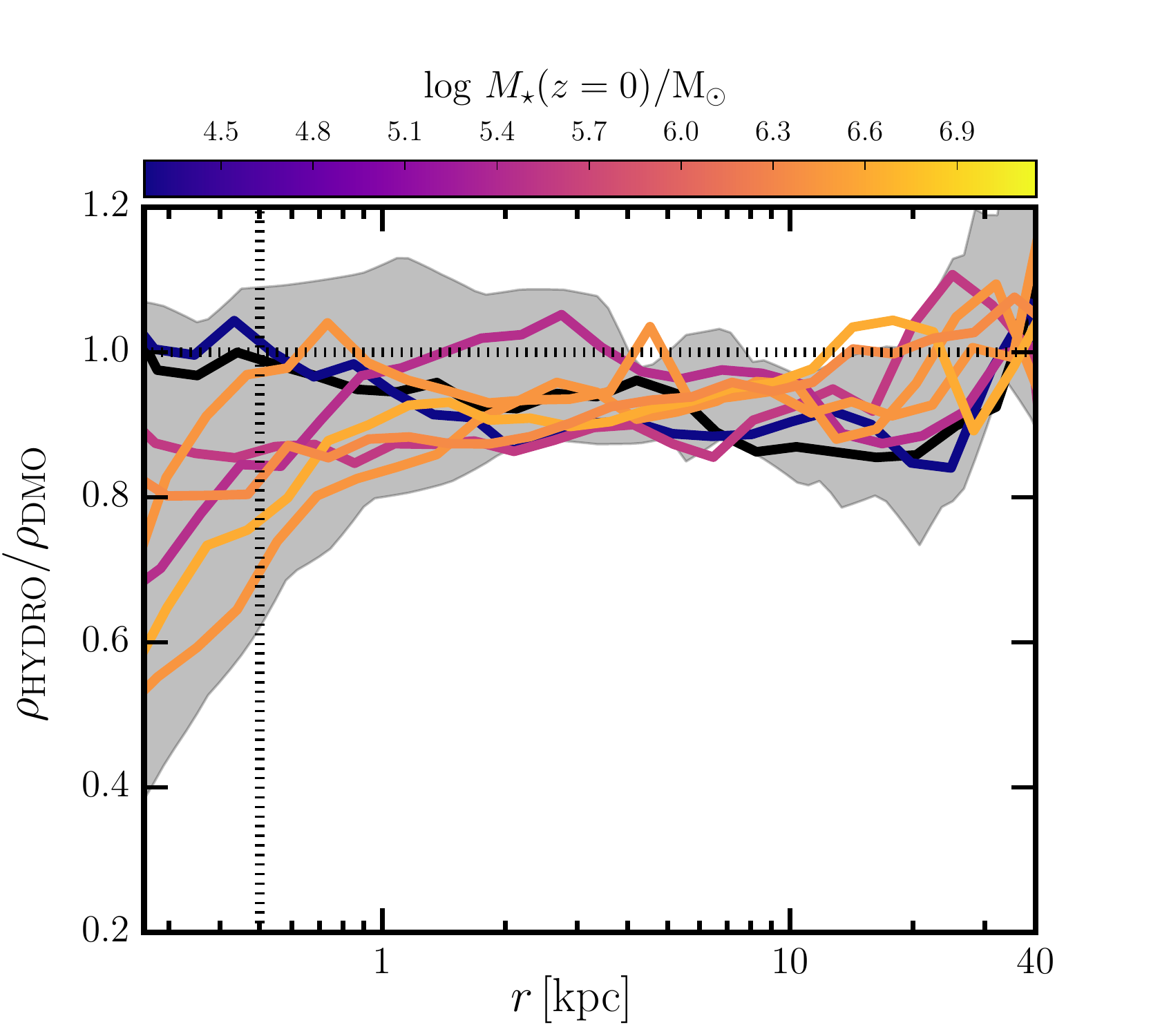}
\includegraphics[width=0.49\textwidth]{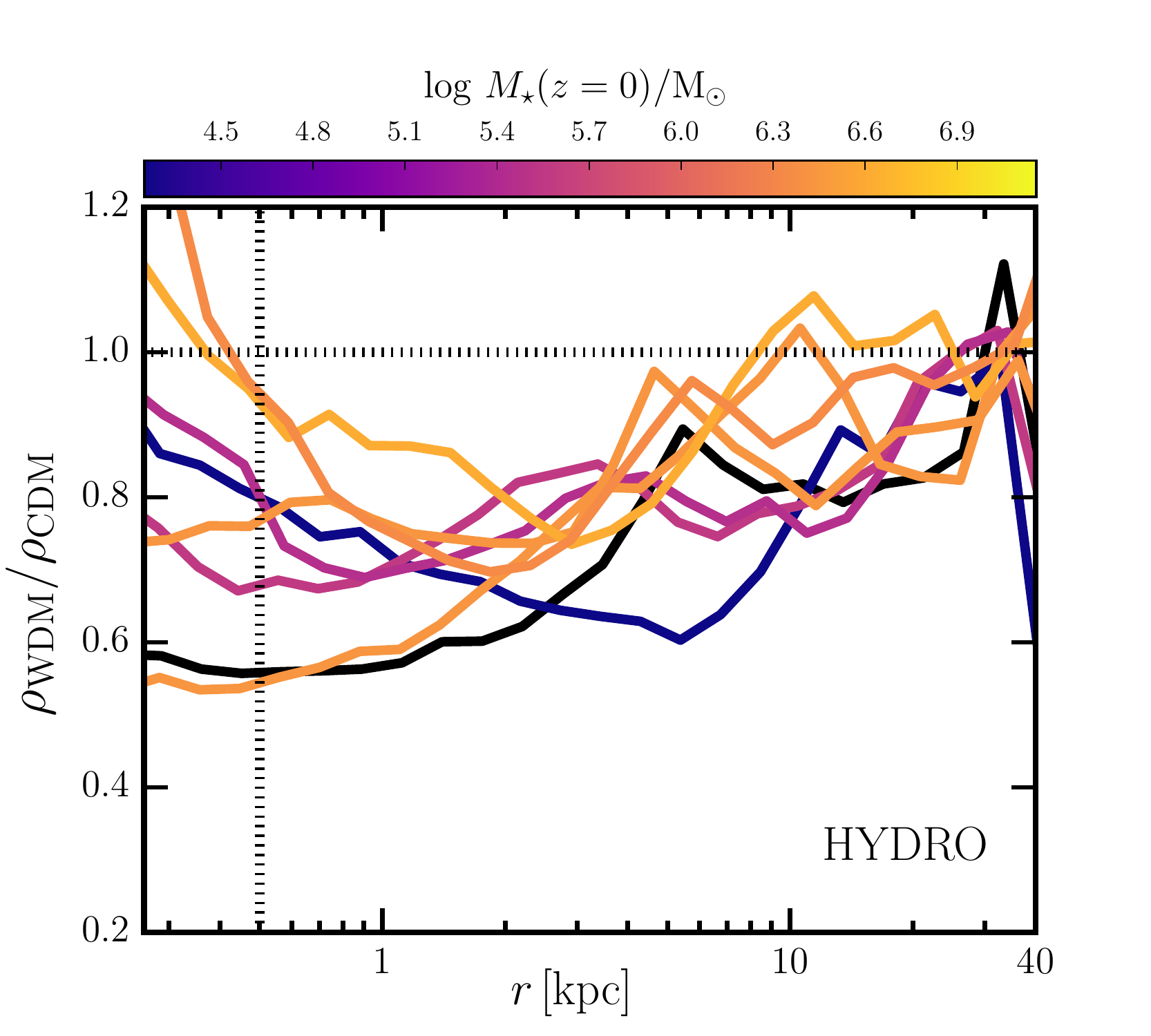} 
\caption{\textit{Left Panel:} The density profile ratios of halos in the hydrodynamical simulations compared to DMO simulations; line styles are the same as in previous figures. To first order, the stellar mass of the galaxy in WDM and CDM is correlated with the suppression of the central dark matter density from stellar feedback effects. Halo m10h, Halo m10m, and Halo m10e illustrate the second order-effects that require slightly more complicated interpretations. Halo m10h is undergoing a recent merger that raises the density ratio to $\sim1$ at $r = 500\,{\rm pc}$ while maintaining the central density reduction at $r<500\,{\rm pc}$. Halo m10m (hosting a more massive galaxy) quenches early ($z\sim3$) and has a similar density ratio at $r=500$ pc as Halo m10e (which hosts a less massive galaxy), which has formed nearly all of its stars since $z\sim1$. \textit{Right Panel:} Ratio of density profiles between WDM and CDM halos from hydrodynamical simulations. There are two galaxies where the CDM halo is less dense than in WDM inside $r \sim400\,{\rm pc}$ (near the numerical convergence radius). All other halos are less centrally dense in WDM independent of the stellar mass of the galaxy. The dashed vertical line in each panel marks 500 pc; trends in the density at this scale are investigated in Figure~\ref{fig:Dens_Ratios_2}.}
\label{fig:Dens_Ratios}
\end{center}
\end{figure*}

\begin{figure}
\begin{center}
\includegraphics[width=0.49\textwidth]{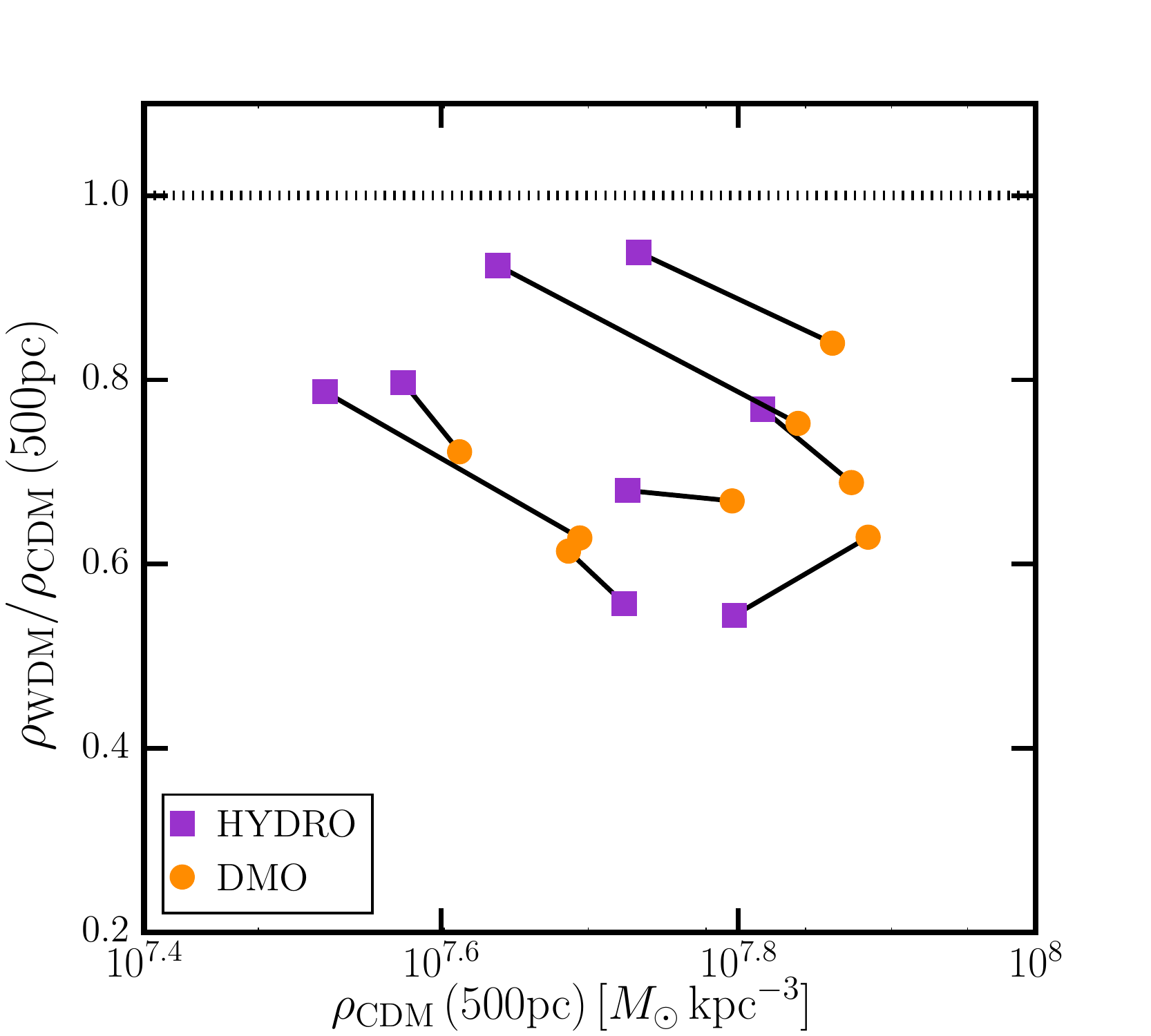} 
\caption{The central density ratio of WDM-to-CDM in DMO (circles) and hydrodynamical (squares) simulations as a function of the density at 500 pc in CDM simulation (DMO or hydrodynamical, as appropriate). The WDM halos are less centrally dense than CDM (at $500\,{\rm pc}$) in both types of simulation. In hydrodynamical simulations, the two most massive CDM galaxies have density reductions that bring them in close agreement with WDM. In nearly all halos, stellar feedback processes are more effective in modifying the central dark matter density in CDM than in WDM, which raises the WDM-to-CDM density ratio for hydrodynamical simulations compared with DMO simulations (in which free-streaming suppresses WDM relative to CDM).}
\label{fig:Dens_Ratios_2}
\end{center}
\end{figure} 

The SFHs of the three remaining WDM halos (Halo m10b, m10c, and m10e) are significantly different than their CDM counterparts; Halo m10b is an extreme example, as it does not form any stars in WDM by $z=0$. Reionization heating suppresses additional accretion of intergalactic gas for Halos m10b and m10c in WDM until $z\sim2$, when the UV background weakens. Halo M10c has a virial temperature at $z_{\rm f}$ that is slightly larger than the temperature to which reionization heats intergalactic gas ($T\approx2\times10^4\,{\rm K}$), resulting in an appreciable reservoir of gas following halo collapse and the ability to accrete gas after $z\sim2$. Halo m10b, on the other hand, has $\tvir < 2\times10^4\,{\rm K}$ for most of its assembly history; it is only at $z\sim0.5$, when it experiences a major merger, that it is able to accrete additional warm gas that begins cooling at $z\sim0.2$. With this late-time build-up of cold gas, there is the possibility that Halo m10b will form stars in the future. In the CDM simulation, Halo m10b has a qualitatively similar halo assembly history, but a larger $\vmax$ at all times, resulting in multiple star formation episodes and multiple galaxy mergers \citep{Fitts2018}. In the larger CDM halo suite of \cite{Fitts2017}, there is one halo (m10a) that does not form any stars; that halo has a similarly suppressed $\tvir({\rm z})$ as the WDM version Halo m10b. A major difference between the ``dark" halo in CDM and WDM is the absence of cold gas in the CDM case compared to a significant amount of cold gas in WDM. 

Finally, we consider Halo m10e, which has a larger gas mass at collapse than Halo m10b and Halo m10c despite having an early $\vmax(z)$ evolutionary track that is similar to Halo m10b. As a result, it has an initial burst of star formation immediately following the rapid collapse phase. A merger at $z\sim2$ raises the halo's $\vmax$ above the threshold for late-time gas accretion described above, which drives the halo's late star formation epoch. To summarize, all WDM halos are able to bind warm gas and eventually build sufficient gas densities that enables cooling processes. Earlier collapse more readily allows for the onset of star formation depending on the depth of the central potential and how rapidly the central potential forms.

The right panel of Figure \ref{fig:Mvir_Mstar} shows the unique SFHs of WDM Halo m10c and Halo m10e, which  form all of their stars after reionization (with the majority of their stellar mass built-up after $z\sim1$). Leo A is the one observed galaxy with a similar SFH featuring a significant late build-up of stellar mass \citep{Cole2007}. However, Leo A has a larger final stellar mass and a non-zero mass fraction of old stars \citep{Cole2014}, both of which clearly distinguish it from the properties of Halo m10c and Halo m10e. Simulations within the CDM paradigm are able to produce strongly delayed SFHs similar to WDM Halo m10c and m10e \citep{Ricotti2009,Shen2014,Fitts2017,Wright2018} but with ancient star formation that better matches Leo A. It is not clear if star formation can be prevented altogether at early times at this mass scale in CDM. The existence of a population of young, gas-rich, actively star-forming ultra-faint dwarfs, including extreme cases like Halo m10c which does not have any stars older than 2~Gyr, makes for a strong prediction of WDM dwarf galaxy formation. 

Other WDM galaxy properties follow CDM galaxy scaling relations. Figure \ref{fig:size_relations} shows the relationship between the stellar half-mass radius of the galaxy, $r_{1/2}$, and the total stellar mass. WDM galaxies are smaller and less massive, but lie along a similar $\mstar-r_{1/2}$ relation as their CDM counterparts. For WDM the relation is given by: $r_{\rm half}=0.51\,\kpc\,(\mstar/10^6 \msun)^{0.365}$. The $\mstar-r_{1/2}$ relationship is also very similar self-interacting dark matter \citep{Robles2017}, as is shown in Figure \ref{fig:size_relations}. However, the shapes of the WDM galaxies and their dark matter halos are triaxial with a similar distribution of axis ratios as their CDM counterparts, which differentiates them from SIDM halos and galaxies \citep{Robles2017}.

\subsection{Feedback Effects on Warm Dark Matter Halo Structure}

In this section, we evaluate the impact of stellar feedback on the structure of dark matter halos. The left panel of Figure \ref{fig:Dens_Ratios} shows the ratio of the density profiles of the WDM halos from the hydrodynamical simulations compared with density profiles from the DMO simulations. The central density of many WDM halos in the hydrodynamical simulations are reduced relative to DMO by as much as $45\%$ near the convergence radius of $r_{\rm conv,max}=265\,{\rm pc}.$  As found for $\mvir \approx 10^{10}\msun$ halos in CDM \citep{Fitts2017}, the degree of reduction on the central density profile of WDM halos roughly correlates with the stellar mass, where low-mass galaxies have little to no effect on their halos and more massive galaxies show significant central density reductions. This trend and the overall spread in the WDM density profile ratio follows the  behavior of CDM halos, as indicated by the grey band in the left panel of Figure \ref{fig:Dens_Ratios}.  

The effects of feedback on the density structure of a galaxy's host halo also depends on the galaxy's SFH. For example, Halo m10m and Halo m10e have similar density ratios (between hydro and DMO in the WDM simulation) at $r=500\,{\rm pc}$ even though Halo m10m hosts a galaxy with 10 times the stellar mass of Halo m10e. At $r=265\,{\rm pc}$, Halo m10e has an additional $\sim15\%$ density reduction compared to Halo m10m. The difference in central density profiles of these two halos can be traced to the very recent star formation of Halo m10e, which forms all of its stars after $z=1$, while Halo m10m self-quenches at early times; this is consistent with the idea that late-time star formation is important for modifying dark matter halo density profiles \citep{Onorbe2015,Read2016}. Merger histories can also play a significant role in determining the central density profiles in the hydrodynamical simulations. For example, Halo m10h hosts one of the more massive WDM galaxies and has a $25\%$ density reduction at $r=265\,{\rm pc}$, similar to other massive galaxies. However, owing to an ongoing merger of a less massive halo, the density ratio (again, between the hydrodynamic and DMO runs in WDM) at $r\sim500\,{\rm pc}$ is close to unity. 

The ratio of density profiles between WDM and CDM from hydrodynamical simulations is shown in the right panel of the Figure \ref{fig:Dens_Ratios}. At $r = 500\,{\rm pc}$, all WDM halos are less dense than their CDM counterparts, and they remain less dense out to $r\gtrsim10\,\kpc$. There is no correlation between WDM galaxy stellar mass and the degree of WDM halo central density reduction relative to CDM, as indicated by comparing the color of the curves in Figure \ref{fig:Dens_Ratios} to the density ratio within the central kpc. For example, Halo m10b has one of the larger central density reductions in WDM relative to CDM in the hydrodynamic simulation even though it does not form any stars in WDM, while Halo m10f hosts one of the more massive WDM galaxies yet has a similar central density profile ratio. Halos m10k and m10m host the two most massive galaxies in CDM and have density ratios $\rho_{\rm WDM}/ \rho_{\rm CDM} > 1$ in the inner $r=400\,{\rm pc}$. These two galaxies have significantly more stellar mass in CDM than in WDM, which results in a greater central density reduction at these radii in CDM. As this is near the convergence radius of our halos, we restrict further analysis to $r = 500\,{\rm pc}$, where our results are more robust.

Figure \ref{fig:Dens_Ratios_2} summarizes the effects of the intrinsic reduction to the central density profiles of WDM halos originating from the delay in halo formation (i.e., coming from the free-streaming length) compared with the effects of feedback in both CDM and WDM hydrodynamical simulations. While galaxies in WDM halos are less massive and experience fewer significant feedback events than in CDM, the combination of late collapse and baryonic processes leave WDM halos less dense at $r=500\,{\rm pc}$ in all cases. The two most massive galaxies in our CDM suite have central dark matter halo densities that only exceed their WDM counterparts by $\sim5-10\%$; all other halos remain $\sim20-45\%$ less centrally dense in WDM relative to CDM.

It is difficult to predict \textit{a priori} whether feedback generally will be more effective at modifying the central density of a CDM halo or its WDM counterpart. On the one hand, the CDM galaxies are typically more massive, meaning they have a larger amount of feedback energy available. On the other hand, WDM halos form with a reduced central density from free-streaming effects, so it might be possible that even diminished feedback input could have a large effect. 
In most of our simulated halos, stellar feedback processes are more effective at modifying the central dark matter density in CDM than they are in WDM, which raises the ratio of WDM central density relative to CDM in Figure~\ref{fig:Dens_Ratios_2}. It therefore appears that the reduced densities of WDM halos coming from free-streaming do \textit{not} generally make these halos more susceptible to further density reduction from stellar feedback. 

However, this is not always the case. In three halos (Halo m10b, Halo m10d and Halo m10f), the ratio decreases or remains flat for differing reasons. Halo m10b has a smaller ratio in the hydrodynamic simulations compared with DMO because the CDM halo in the hydro simulation is denser than it is in DMO, while the central density in the WDM DMO and hydro simulations are identical. The density ratio for Halo m10d is nearly flat; this is because the reduction in the central density profile by stellar feedback is comparable in WDM and CDM (despite the smaller stellar mass and abbreviated SFH in WDM). Halo m10f is the one WDM halo where feedback effects are \textit{more} effective at reducing the central density profile in WDM than in CDM. The extra reduction is modest ($\sim10\%$); however, this shows that it is possible to see a larger modification of the density profile in WDM relative to CDM in some cases. Of all the halos in our suite, Halo m10f has the most similar stellar mass and average star formation rate in both WDM and CDM cosmologies, which may account for the effectiveness of stellar feedback in modifying its central density in WDM.

\section{Discussion}
\label{sec:Disc}

\subsection{Predicting trends with WDM Halo Mass and WDM Model Parameters}

A central motivation of this work is to understand the degree to which the combination of stellar feedback effects and dark matter physics modifies the dark matter distribution in WDM halos relative to CDM halos. We have shown that these combined effects make WDM halos less centrally dense than their CDM counterparts. In this section, we consider extrapolation of those results to different WDM models and different halo masses, as results of the previous sections are specific to $\mvir(z=0)=10^{10}\,\msun$ halo mass scale in our simulation suite and the free-streaming scale of the S229 resonantly-produced sterile neutrino WDM model. 

Assembly histories of a $\mhalo = 10^{10}\,\msun$ in a colder model (e.g. the S220 resonantly-produced sterile neutrino WDM model discussed in \citealt{Bozek2016}) would approach CDM assembly histories. S220 dark matter halos would collapse earlier and form with a larger central density (or equivalently larger $\vmax(z_{\rm f})$) than the S229 model halos but with a lower central density than CDM. Given that the slope of the $\vmax(z_{\rm f})-M_{\star}(z=0)$ in the left panel of Figure~\ref{fig:VmaxMstar} is similar for CDM and S229, a colder model should follow the same trend with points falling between the S229 and CDM values. The central galaxies would have larger stellar masses, and consequently provide more energy from stellar feedback, but would form initially in a denser dark matter halo than in our WDM case. 

An interesting topic for future work is to determine whether the greater amount of stellar feedback in a denser $\mhalo = 10^{10}\,\msun$ halo of a colder WDM model will prove more or less efficient at modifying the central dark matter density and result in a smaller or larger $\rho_{\rm WDM}/ \rho_{\rm CDM}(r=500\,{\rm pc})$ ratio. However, we note that Figure \ref{fig:Dens_Ratios_2} shows that several of our WDM halos are able to match or exceed the overall reduction in central density compared with CDM despite having smaller stellar masses. It is plausible that there may be a WDM model parameterization (between S229 and CDM) with a maximal reduction in the density profile relative to CDM that exceeds what we find in the right panel of Figure \ref{fig:Dens_Ratios}. 

While only one ultrafaint dwarf ($\mstar \lesssim 10^{5}\,\msun$) in our WDM simulations is formed with a small fraction of ancient (pre-reonization) stars, our results do not preclude the possibility that the S229 WDM model studied here can produce significant numbers of ultrafaint dwarfs with predominantly ancient stellar populations. However, the delay in formation time for the S229 model will limit the fraction of WDM halos at this mass that will collapse early enough to form old stars (see also \citealt{Governato2015}). Lower-mass halos will suffer a longer delay in forming \citep{Bose2016}, leaving only higher-mass WDM halos that quench early in their assembly history (similar to CDM models) to additionally host old ultrafaint dwarfs in WDM.

It is difficult to cleanly extrapolate from our results, which are based on galaxies forming in isolated environments, to counts of galaxies in and around the Local Group, where the large-scale overdensity  may lead to earlier collapse times and earlier star formation. However, a direct comparison with subahlo $\vmax$ functions of Local Group analogues does point to difficulties in matching ultrafaint counts in the S229 WDM model if ultrafaints are all predominantly old. Cooler WDM models will have larger fractions of old stellar populations and may not have a similar limitation in their assembly histories. Colder models are also likely to have a population of dark (starless) halos, similar to CDM \citep{Sawala2013,Benitez-Llambay2017,Fitts2017}.

\subsection{Implications for reionization in WDM}

Half of the WDM dwarf galaxies simulated here do not form any stars before the end of reionization ($z=6$). The galaxies that do form stars prior to $z=6$ have no star formation prior to $z=7$, however, so they contribute minimally to the ionizing photon background. Their CDM counterparts all form a significant mass of stars prior to the end of reionization. The brightest CDM galaxies in our suite have $\mstar \sim 10^6\,\msun$ at $z=6$. Extrapolating the stellar mass-UV luminosity relation from \citet{Ma2017}, these galaxies will have a UV luminosity of ${\rm M_{\rm UV}} \sim -12$.

In the CDM paradigm, these galaxies make-up a significant fraction of the galaxy population that drives cosmic reionization (e.g., \citealt{Livermore2017} and references therein). The galaxy population that sources reionization in WDM is more uncertain. Semi-analytic modeling and abundance matching techniques applied to large-volume cosmological simulations adopting  WDM models of comparable warmth to the S229 model demonstrate that reionization can be completed by $z=6$ while matching constraints on the epoch of reionization from \textit{Planck} \citep{Schultz2014,Bose2016b}. However, the abundance of low-mass halos at high redshift in WDM models is significantly reduced. For example, the S229 model only has about $40\%$ as many halos with $\vmax(z=6) \sim 35\,\kms$ relative to CDM \citep{Bozek2016}. 

As a result, galaxies in more massive halos must drive reionization in the WDM paradigm, possibly with a different SFH and ionizing budget than what is predicted for CDM \citep{Bozek2015, Bose2016b,Villanueva-Domingo2018}. Alternatively, other sources such as quasars could be a more important contributor in WDM cosmologies. Including additional sources or considering a different UV background (either in intensity or time) than the one used here could possibly further suppress the late-time SFHs of the low-mass galaxies. Future probes of reionization and counts of galaxies at high redshift provide a important opportunity to constrain WDM models.

\subsection{Predicted galaxy populations in WDM}

There are features in the galaxy populations that can be used to distinguish between dark matter models. The WDM halos at this mass scale do not form stars prior to $z=7$. This makes it difficult to account for observations of ultrafaint galaxies in the Milky Way that quench early and have a population entirely made up of old stars \citep{Brown2014,Weisz2014}.\footnote{This assumes that the truncation in the ultrafaint dwarf SFH is set by reionization feedback effects. Recent observations of M31 ultrafaint dwarf galaxies have found extended SFHs, implying that the Milky Way utlrafaint satellite SFHs could be a product of local environmental effects such as non-uniform patchy reionization or stripping by the host \citep{Martin2107}.} Extrapolating the stellar mass -- halo mass relation informed by simulations in the CDM paradigm, ultrafaint galaxies are expected to be hosted in $\mhalo \approx 10^{9} \,\msun$ halos \citep{Moster2013,Munshi2013,Munshi2017}. As discussed above, isolated ultrafaint galaxies with a significant fraction of stars formed prior to the end of reionization might be hosted by WDM halos with a range of halo masses, including larger masses ($\mhalo > 10^{10}\,\msun$) than the halos simulated here. This would introduce significant scatter in the low-mass end of the stellar mass -- halo mass relation, which has also been proposed in CDM cosmologies \citep{Garrison-Kimmel2017a}. 

With a fixed number of $10^{10} \,\msun$ WDM halos and a different stellar mass -- halo mass relation than in CDM, the faint-end slope of a WDM stellar mass function of field galaxies would be flatter than would be naively expected from halo counts alone predicted by DMO simulations in CDM \citep{Bozek2016}. Counts of faint galaxies detected in future imaging and HI surveys are therefore likely to provide strong constraints on WDM models.

A unique prediction of our suite of WDM galaxies is a population of ultrafaint dwarf galaxies that are extremely young, forming over $80\%$ of their stars in the last four Gyr. These galaxies are gas-rich and actively star forming at $z=0$. Such SFHs are not entirely unlike Leo A \citep{Skillman2014}; however, neither of our late-blooming galaxies has \textit{any} star formation prior to $z\approx3.5$, unlike Leo A \citep{Cole2007}. There is one galaxy in the CDM sample that forms over 80\% of its stars in the last two~Gyr \citep{Fitts2017}, but it also has a population of ancient ($z\sim7$) stars, in contrast with the young WDM cases. 

Both dark matter models predict halos that do not form any stars, so it is possible that a CDM model could produce a similarly young ultrafaint dwarf. In that case, the central density profile of the dwarf could help to constrain which type of dark matter halo it resides in, as it would be at least $20\%$ denser in CDM than in WDM according to Figure \ref{fig:Dens_Ratios_2}. Further, if similarly young galaxies form in CDM, they would likely be less abundant. The discovery of a population of galaxies with entirely young stars residing in halos at this mass scale would therefore strongly disfavor CDM and favor a WDM model.

\section{Conclusions} 
\label{sec:Conc}

In this paper, we have presented simulations of eight isolated dwarf galaxies with a halo mass of $\mhalo = 10^{10}\,\msun$ in WDM using the FIRE-2 galaxy formation model. The underlying dark matter particle model is a resonantly-produced sterile neutrino with a mass of $m_{\rm s} = 7.1\,{\rm keV}$ and a mixing angle of $\sin^{2}(2\theta) = 2.9 \times 10^{-11}$. This model was selected to (1) provide free-streaming effects that are at the warm edge of what is allowed based on satellite galaxy counts and large-scale structure constraints and (2) account for the origin of possible detections of an X-ray line at $3.55\,{\rm keV}$ in galaxy and galaxy cluster observations. These isolated dwarfs were also studied in CDM in \citet{Fitts2017}. 
A summary of our main results is as follows:
\begin{itemize}

\item WDM halos collapse later and with a larger initial mass than their CDM counterparts. This results in a shorter, near-monolithic period of initial collapse that allows WDM halos to catch up to their CDM counterparts in the halo assembly process. Following this rapid collapse phase, the central halo potential (as measured by the halo $\vmax$) is set at a similar cosmic period of $t\sim2-4\,{\rm Gyr}$ in both CDM and WDM. WDM free-streaming effects weaken density perturbations on small scales, resulting in WDM halos forming with a decreased central density, or equivalently, a smaller peak $\vmax$. The reduction in WDM central density relative to CDM is approximately the same for all halos with a density ratio ranging from $\rho_{\rm WDM}/\rho_{\rm CDM}(r=500 \,{\rm pc}) = 0.60-0.75$ and a mean value of $0.70$. 

\item The stellar mass of a galaxy at $z = 0$ is correlated with its host halo's central density (or $\vmax (z_{\rm f})$) at the time of formation, where $z_{\rm f}$ is defined as the redshift where the $\vmax$ function initially reaches a value of $0.85\, \vmax(z=0)$. The WDM halos follow the same $\mstar(z=0)-\vmax(z_{\rm f})$ trend as the CDM halos owing to a similar degree of central density reduction from free-streaming effects amongst our WDM halos.

\item Our suite of WDM dwarf galaxies can be divided into two groups based on stellar mass: WDM halos with $\vmax(z_{\rm f}) > 25\,\kms$ all have $\mstar \gtrsim 2\times10^6\,\msun$. WDM halos with $\vmax$ below this threshold form far fewer or no stars. These halos are strongly affected by reionization feedback, as they collapse after $z=6$ or experience significant late-time quenching events and have shallow central potentials that are minimally sufficient in accreting and cooling warm ($T_{\rm IGM} \sim 2\times10^4\,{\rm K}$) gas in order to form stars.

\item The galaxy scaling relations follow the same trends in WDM as in CDM. Given a galaxy of a fixed mass, shape, and size, it is not possible to identify whether the halo is made up of warm or cold dark matter. All WDM galaxies feature a delay in the onset of star formation due to the later halo collapse times. As a result, all of the WDM galaxies studied here are devoid of stars formed before  $z=7$ and contain a significantly smaller fraction of stellar mass formed at early times relative to their CDM counterparts. 

While our WDM simulation suite includes an ultrfafaint galaxy with some ancient (reionization-era) star formation, there are no ultrafaint galaxies with an entirely old stellar population in this suite. Moreover, we predict a population of extremely young, gas rich, actively star forming ultrafaint dwarfs that have formed over $80\%$ of their stars in the last four~Gyr if dark matter is warm. Both WDM and CDM predict that some $\mhalo = 10^{10}\,\msun$ halos will not be able to form any stars. 

\item  Free-streaming reduces the central densities of dwarf WDM halos relative to their CDM counterparts. Feedback from bursty star formation further reduces the central density of WDM halos; this effect is also present in CDM simulations. The CDM analogs host more massive galaxies and have more significant feedback events; the CDM galaxies therefore tend to have a greater impact on their halos' central densities compared to their WDM counterparts when isolating the effects of stellar feedback from free-streaming. In one case, however, stellar feedback is more effective at reducing the central density in WDM than in CDM despite a smaller stellar mass and abbreviated SFH for the WDM galaxy. While the extra reduction is a modest $\sim10\%$, this demonstrates that larger modification of WDM halos relative to CDM is possible in hydrodynamical simulations for WDM halos with extended SFHs and low central densities.

\item Only half of our suite of WDM galaxies form stars before $z=6$ and none of the halos form stars before $z=7$. These WDM galaxies contribute negligibly to the ionizing photon background, contrary to the expectations for galaxies in halos at this mass scale in CDM. Previous studies have shown that WDM models with a similar free-streaming scale as the S229 model we study here are able to complete reionization by $z=6$ through vigorous star formation in halos with $\mhalo > 10^{10}\,\msun$ \citep{Schultz2014,Bose2016b,Villanueva-Domingo2018}. The ability of these models to match high-z UV luminosity functions and the implications for galaxy formation in larger mass WDM halos is left for future work.

\end{itemize}

We have demonstrated the viability of a resonantly-produced sterile neutrino WDM model to produce realistic dwarf galaxies and to simultaneously address small-scale issues found in the CDM paradigm. Our suite of WDM halos features density profiles that are less centrally dense than their CDM counterparts, which is important for addressing the TBTF problem. A colder model and/or higher-mass halos may see even larger reductions in the central density. A colder resonantly-produced sterile neutrino WDM model may also provide a better fit to constraints from reionization and SFHs of Local Group ultra-faint dwarfs. We will consider such models through FIRE simulations in a future paper.

\section{Acknowledgments}
MBK and AF acknowledge support from NSF grant AST-1517226. MBK was also partially supported by NASA grants NNX17AG29G and HST-AR-13888, HST-AR-13896, HST-AR-14282, HST-AR-14554, HST-GO-12914, and HST-GO-14191 from the Space Telescope Science Institute, which is operated by AURA, Inc., under NASA contract NAS5-26555. KNA is supported by NSF Theoretical Physics Grant No. PHY-1620638. DK was supported by NSF grant AST-1715101 and the Cottrell Scholar Award from the Research Corporation for Science Advancement. Support for SGK was provided by NASA through Einstein Postdoctoral Fellowship grant number PF5-160136 awarded by the Chandra X-ray Center, which is operated by the Smithsonian Astrophysical Observatory for NASA under contract NAS8-03060. CAFG was supported by NSF through grants AST-1412836, AST-1517491, AST-1715216, and CAREER award AST-1652522, by NASA through grant NNX15AB22G, and by a Cottrell Scholar Award from the Research Corporation for Science Advancement.

\bibliography{DM_heats_up.bib}
\bibliographystyle{mnras.bst}
\label{lastpage}
\end{document}